\documentclass{article}
\usepackage{arxiv}
\usepackage[utf8]{inputenc} 
\usepackage[T1]{fontenc}    
\usepackage{hyperref}       
\hypersetup{
    colorlinks = true,
    citecolor = {blue},
}
\usepackage{url}            
\usepackage{booktabs}       
\usepackage{amsfonts}       
\usepackage{nicefrac}       
\usepackage{microtype}      
\usepackage{graphicx}
\usepackage{caption}
\usepackage{subcaption}
\usepackage{natbib}
\usepackage{hypernat}
\usepackage{adjustbox}
\usepackage{amsmath,bbm}
\usepackage{graphicx,psfrag,epsf}
\usepackage{enumerate}
\usepackage{amsthm,amssymb}
\usepackage{mathtools}
\usepackage{xcolor}

\usepackage[inline]{trackchanges}
\addeditor{LS}
\addeditor{LP}
\addeditor{JDT}
\addeditor{GH}

\usepackage{url} 
\usepackage{algpseudocode}

\newcommand{\bomega}{\boldsymbol{\omega}}

\newcommand{\btheta}{\boldsymbol{\theta}}
\newcommand{\brho}{\boldsymbol{\rho}}

\newcommand{\N}{\mathcal{N}}
\newcommand{\bx}{\mathbf{x}}
\newcommand{\bu}{\mathbf{u}}
\newcommand{\by}{\mathbf{y}}

\newcommand{\bs}{\mathbf{s}}
\newcommand{\bc}{\mathbf{c}}

\newcommand{\bm}{\mathbf{m}}

\newcommand{\dif}{\mathrm{d}}
\newcommand{\E}{\mathbb{E}}

\newcommand{\hist}{\mathcal{H}_t}
\newcommand{\X}{\mathcal{X}}
\newcommand{\Z}{\mathcal{Z}}
\newcommand{\B}{\mathcal{B}}
\newcommand{\D}{\mathcal{D}}

\title{Spatio-temporal extreme event modeling of terror insurgencies}

\author{
  Lekha~Patel\thanks{Corresponding author: \texttt{lpatel@sandia.gov}} \\
  \And
  Lyndsay Shand \\
  \And
  J. Derek Tucker \\
  \And
  Gabriel Huerta \\
  \And
  \vspace{-10pt} \\
  Statistical Sciences, Sandia National Laboratories, Albuquerque, NM \\
}

\begin{document}
\maketitle

\begin{abstract}
Extreme events with potential deadly outcomes, such as those organized by terror groups, are highly unpredictable in nature and an imminent threat to society. In particular, quantifying the likelihood of a terror attack occurring in an arbitrary space-time region and its relative societal risk, would facilitate informed measures that would strengthen national security.
This paper introduces a novel self-exciting marked spatio-temporal model for attacks whose inhomogeneous baseline intensity is written as a function of covariates . Its triggering intensity is succinctly modeled with a Gaussian Process prior distribution  to flexibly capture intricate spatio-temporal dependencies between an arbitrary attack and previous terror events. By inferring the parameters of this model, we highlight specific space-time areas in which attacks are likely to occur. Furthermore, by measuring the outcome of an attack in terms of the number of casualties it produces, we introduce a novel mixture distribution for the number of casualties. This distribution flexibly handles low and high number of casualties and the discrete nature of the data through a {\it Generalized ZipF} distribution. We rely on a customized Markov chain Monte Carlo (MCMC) method to estimate the model parameters. We illustrate the methodology with data from the open source Global Terrorism Database (GTD) that correspond to attacks in Afghanistan from 2013-2018.  We show that our model is able to predict the intensity of future attacks for 2019-2021 while considering various covariates of interest such as population density, number of regional languages spoken, and the density of population supporting the opposing government.
\end{abstract}


\section{Introduction} 
\label{sec:introduction}
Recent global developments have seen a surge in high-consequence events with life-threatening consequences. The emergence of such extreme situations has heightened the urgency of their detection, prevention, and deterrence. While many quantitative methods endeavor to address this topic, few are able to fully consider the spatio-temporal impact of historic records on the risk of subsequent lethal events. In this paper, we propose a novel approach to detecting and predicting terrorism attacks using concepts from extreme value analysis and a spatio-temporal self-exciting point process framework. The focus of this work is in modeling \textit{terror insurgencies}, violent uprisings organized by groups such as the Taliban, Boko Haram, and the Islamic State (IS), using a repertoire of terror-based attacks against civilians in an attempt to overthrow a government of interest. Application of our model will be highlighted by studying the insurgency of the Taliban in Afghanistan between 2013-2021, a highly relevant socio-political topic that is currently generating significant global coverage.

In the literature, traditional statistical models seek to describe the overall distribution of such rare events of modeling interest, while extreme value analysis prioritizes the characterization of events that lie in the tails of these distributions. Extreme value methods have commonly been used to predict and quantify uncertainty around environmental or climatological events associated with high costs such as a high impact on human casualties (e.g., earthquakes, hurricanes, flooding, wildfires), as described in for example, \cite{RN218}. 
From a statistical point of view, extreme value data can be modeled based on two types of approaches. The first relies on calculating a sequence of maximum (or minimum) values over blocks of data, e.g. monthly or yearly maxima (minima) and fitting these values to their large sample distribution, the Generalized extreme value (GEV) distribution. The second  finds observations that exceed (or fall below) a given threshold and fits the Generalized Pareto distribution (GPD) to these exceedance values. The latter method is also known as a Peaks over Threshold (PoT) approach and can make a more effective use of the data rather than only considering the block maxima. Additionally, a PoT approach can be more flexible than analyses of block maxima as it can allow for the simultaneous prediction of the time and characteristics of an extreme event. The utilization of both methods are application specific but have not been considered extensively to assess extreme high consequence terrorism events. However, the specialized literature includes extensive developments of both block-maxima and PoT approaches for time series and point-referenced spatial data. The interested reader is directed to the works of \cite{RN189, RN225, RN227, RN192, RN191, RN203, RN228, RN217, RN230} for more information on temporal or space-time extreme value analysis and some of its applications. 

In this work, we propose novel methodology to analyze spatio-temporal patterns of extreme terror attacks with potential life-threatening implications. Our model is intended to quantify and predict the time, location, and likelihood of a future extreme event, with the extremity of an event defined by the number of casualties it produces. A specific aspect of our model for this application considers point process modeling and therefore follows the works of \cite{patel2020,tucker:19,RN597}. Point process models are useful to describe, in a probabilistic manner, phenomena that occur in space-time, and are therefore highly suitable to model complex terror insurgencies. In such situations, {\it homogeneous} point processes, which would assume a constant rate of attacks, insufficiently capture event clustering in space-time. Instead, we rely on self-exciting processes to capture the spatio-temporal dependencies that naturally arise in terror formations and resurgences, as illustrated previously by \cite{RN593} which focused on crime modeling and dispersion in Chicago and \cite{tucker:19} which looked at issues of incomplete data to characterize terrorism events in Colombia. 

Our methodology is sufficiently flexible to characterize the spatio-temporal behavior of region-specific attacks and their associated number of casualties, which represent the marks of a marked spatio-temporal point process underlying terror behavior. To account for the extreme behavior of these marks, we develop a novel mixture model to estimate a discrete \textit{mark} distribution that simultaneously models situations with a low number of casualties and cases where casualties can be considered extreme or \textit{above a threshold}. In particular, we show that the marks above a threshold are well estimated via a {\it Generalized-ZIPF} distribution which gives an analog representation to the well known GPD or PoT approach as described previously. Therefore, we can appropriately model the number of casualties for extreme terrorism events separately from events which fall within the expected casualty range. In the context of financial time series, \cite{GROTHE2014269} proposed a multivariate time series model within the PoT framework where clustering of events is captured via a self exciting point process.  The model parameters are estimated with MLE and  illustrated with the analysis of  daily log-returns obtained from financial markets.   To the best of our knowledge, the combination of self-exciting mark point processes with a discrete extreme value distribution has not been considered before for the analysis and prediction of terror insurgencies.  

This paper is organized as follows. Section \ref{sec:gtd} introduces our motivating dataset, the Global Terrorism Database previously considered by \cite{tucker:19} and the different aspect of focus for this work. Section \ref{sec:methods} presents our novel and flexible marked point process framework and details the mark distribution choices made via a mixture representation. It also discusses an inference approach based upon a novel application of Markov chain Monte Carlo (MCMC) methods. Section \ref{sec:results} applies our model to data from the GTD and in difference to \cite{tucker:19} we focus on data from 2013-2018 corresponding to Afghanistan and illustrate the capability of our model in predicting the intensity of future events. Section \ref{sec:disc} gives our conclusions and discusses potential extensions of our work. 

\section{Data}
\subsection{Global Terrorism Database} 
\label{sec:gtd}

The methodology proposed in this paper focuses on the Global Terrorism Database, \cite{GTD}.  The GTD is an open-source database which can be downloaded at \url{https://www.start.umd.edu/gtd/}, and includes information on more than 190,000 terrorist attacks from 199 different countries occurring from 1970 to 2018. Along with the attack time (day, month and year) and location (latitude and longitude) of each recorded terrorism event, detail is also given on the attack type, weapon(s) used, nature of the target, casualties, injuries, and the group responsible (when available). While extreme value theory has been shown to be a valid approach for predicting the likelihood of catastrophic terrorism events in \cite{RN201}, using a more limited data set than the GTD, these analyses had not been extended to appropriately incorporate spatial or other critical descriptive information of the terrorism events. 

\begin{figure}[htbp]
	\centering
	\begin{subfigure}{.45\textwidth}
	  \centering
	  \includegraphics[height=7.7cm,width=8.6cm]{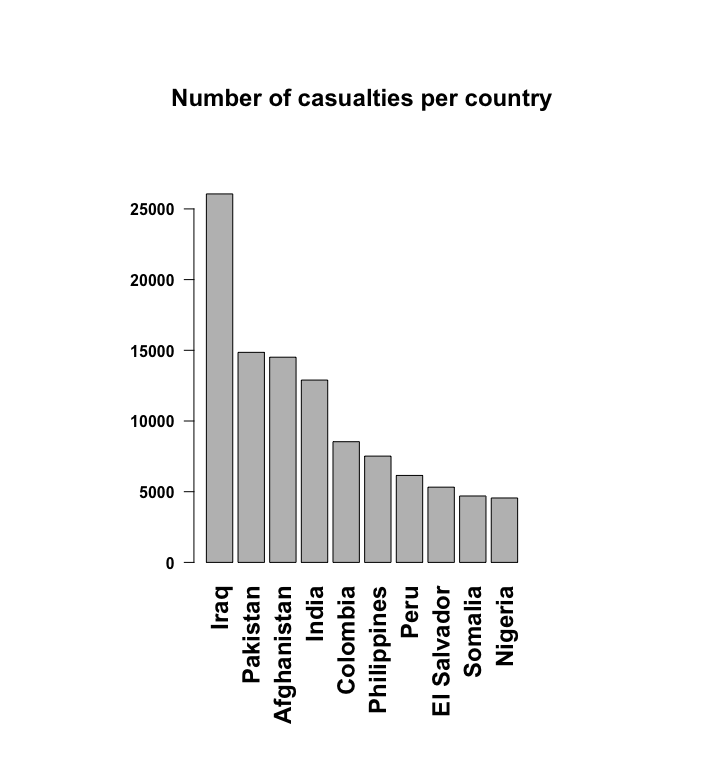} 
	  \caption{Histogram of all casualties from countries with the highest casualties across all attacks since 1970.}
	  \label{fig:hist_all}
	\end{subfigure}%
	\begin{subfigure}{.45\textwidth}
	  \centering
	  \includegraphics[scale=.25]{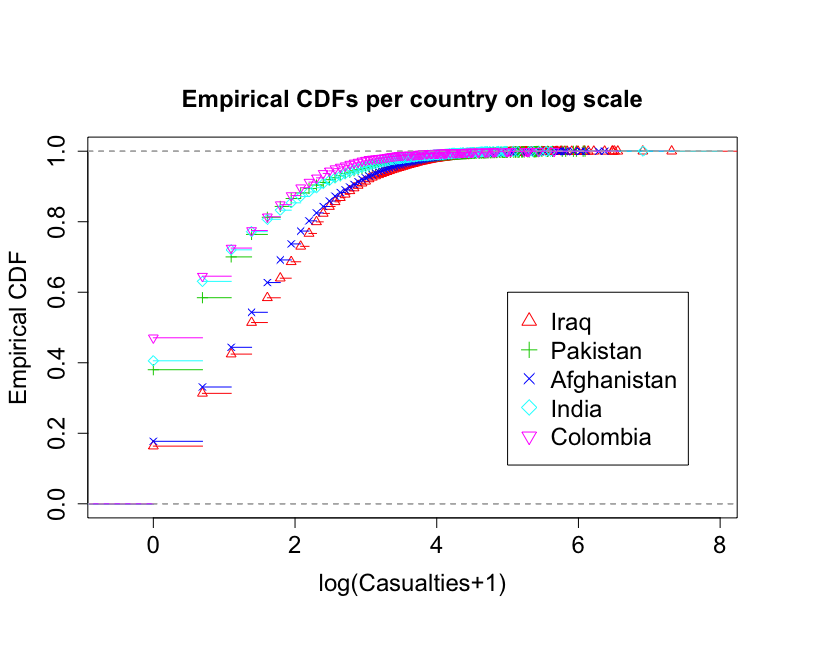} 
	  \caption{Empirical CDF.}
	  \label{fig:emp_cdf}
	\end{subfigure}%
	\hfill \hfill
	\begin{subfigure}{.55\textwidth}
	  \centering
	  \includegraphics[scale=.24]{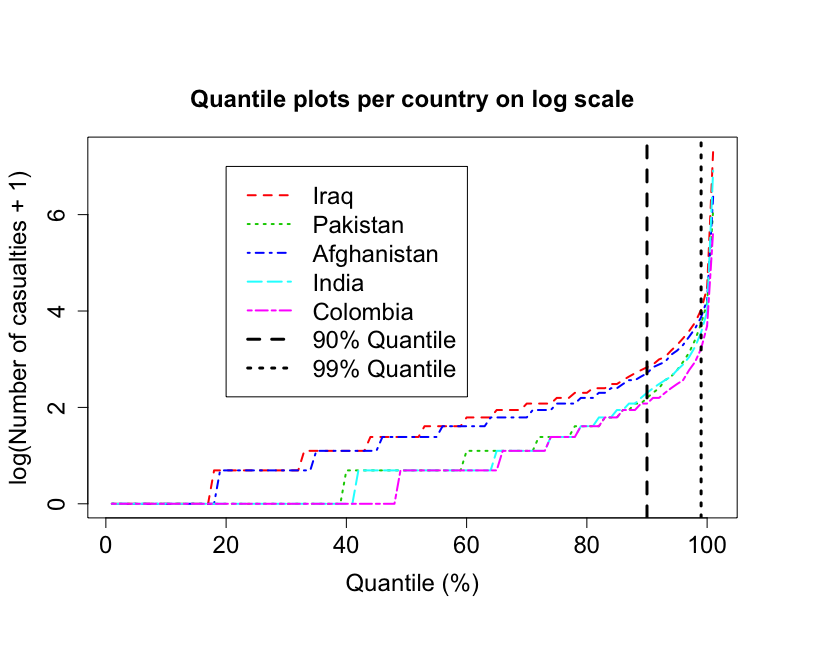} 
	  \caption{Quantile function.}
	  \label{fig:quantile_function}
	\end{subfigure}
	\caption{Plots of all global casualties during 1970-2018.}
	\label{fig:plots_all}
\end{figure}

Although the GTD contains numerous accounts of attacks, events are not consistently distributed evenly across the globe or across time. Data limitations for some countries e.g. the USA and UK from which there is insufficient data, additionally prove challenging for extreme value analysis (EVA).
This therefore limits our focus to specific regions in the world (e.g. countries) where more events occur and thus the resulting analyses can provide meaningful interpretations. The thresholds which define extreme events will be region dependent as seen in Figure \ref{fig:hist_all} which shows the number of casualties per country for cases which reported the highest casualties due to terrorism events since 1970. 

Data quality may also affect subsequent analyses of extreme events. For example, the GTD is incomplete and contains missing entries on one or more of the \textit{explanatory covariates} and \textit{time stamps} of events. The \textit{spatial} locations of events, given by latitude and longitude, are estimated at the nearest city and thus can be inaccurate. As highlighted by \cite{tucker:19}, these aspects of the data are important to account for when developing a statistical model. The data for the year 1993 is completely missing and although there have been efforts to recollect this data, these have not been successful. Additionally, it is possible that many countries, especially those with attacks recorded before 2000, have \textit{underreported} the severity of the attack. Including this data, therefore, would likely lead to misleading results. Although the \cite{GTD} team has made improvements of the methodology used to compile the database, by balancing the strengths of ``artificial and human intelligence'' based on more diverse sets of news media from around the world, we choose to omit this data from our analysis and focus on data from 2013 to 2018.
 
In order to develop an appropriate statistical framework for extreme global terror insurgency attacks, we first require an extreme event to be defined and assess the viability of these distributions to fit the data. Depending on the question of interest, an extreme attack could either be characterized as occurring in an unusual or unexpected space-time location, or as an extremely rare type such as an event perpetrated by a new terrorist group or an event reporting a high magnitude of fatalities or causalities. The former of the two requires a further definition of what constitutes as an ``unexpected'' space-time location and could therefore be difficult to formally formulate in comparison to the latter. We therefore turn to rare type modeling and begin, in a similar manner to \cite{RN201}, by defining an extreme attack as one that results in a high reporting of casualties. The number of casualties here is calculated as the number of deaths and injuries resulting from an attack. 

\begin{figure}
	\centering
  	\includegraphics[scale=.15]{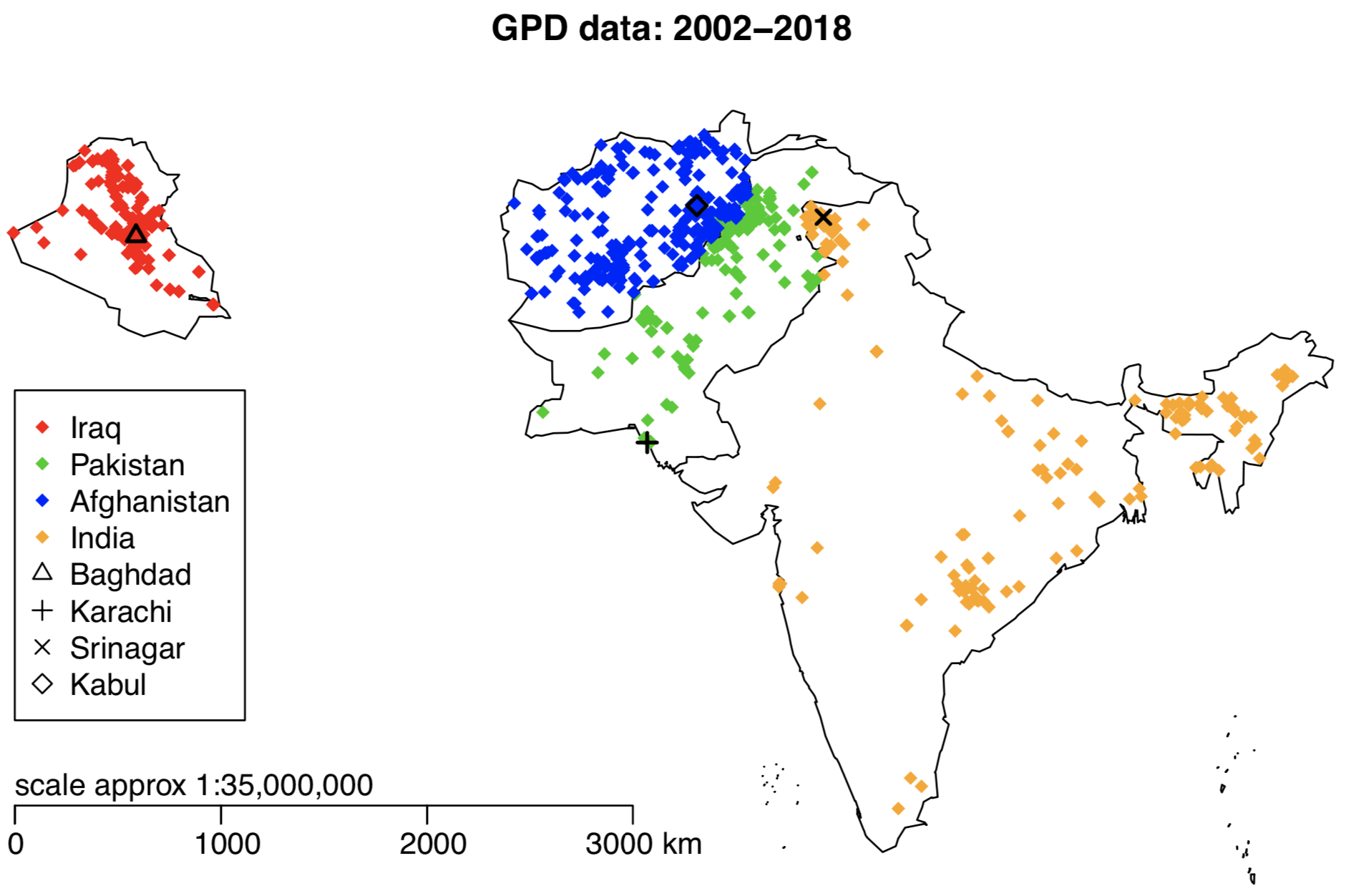}
  	\caption{PoT casualty data between 2002-2018. The locations shown corresponding to cases where the per-country casualty level is above the empirical 97\% quantile}
	\label{fig:cluster}
\end{figure}

On a country specific level, the countries which have sufficient data and quoted the most deadly attacks in the GTD are: Iraq, Pakistan, Afghanistan, India and Colombia as shown in Figure \ref{fig:hist_all}.
Log scaled empirical cumulative distribution functions (CDFs) and quantile plots shown in Figures \ref{fig:emp_cdf} and \ref{fig:quantile_function} respectively, suggest that the distributions of casualties reported from the most deadly countries (Iraq/Afghanistan, India/Pakistan/Colombia) are similar and heavy tailed. These exploratory plots indicate that extreme valued type of distributions would be applicable for this data. 
Figure \ref{fig:cluster} shows the location of terrorist attacks across the region where the per-country number of casualties is above the 97\% empirical quantile. The patterns observed in this figure give an indication that there is a significant spatial contribution on the underlying distribution of extreme events in India, Afghanistan, Iraq and Pakistan, also rendering that high-casualty events lead to clustered behavior.

\section{Materials and methods}\label{sec:methods}
The temporal nature of terrorism events can be identified as being \textit{self-exciting} in nature as described by \citep{RN593,RN592,RN233,tucker:19}, indicating that events occurring in time are likely to be dependent on any previous event in the history of the process. It is also understood that spatially, violent crime activity is likely to occur in or around locations that have had previously successful attacks \citep{RN593}. Such behavior therefore induces clustering patterns and Figure \ref{fig:cluster} shows similar clustering for high causality events. On the other hand, new terrorist activity often arises in regions with denser populations \citep{RN597}. Other variables such as gross domestic product (GDP), ethnicity distribution and geographical terrain additionally indicate regions that are prone to terror events and could be considered as explanatory variables. 

To address the nuanced spatio-temporal behavior of terror events and the casualties that are reported from them, we utilize a marked spatio-temporal modeling approach. This model considers a terror event as an attack taking place at an arbitrary space-time location such that the \textit{mark} of each event defined the number of casualties that occur from it. 

Let $\{X(\bs,t,m): \bs \in W, t \in [0,T], m \in \mathbb{N}_0\}$ denote the marked point process of terror attacks indexed by the spatial location $\bs$ observed in a bounded region $W$, the continuous time $t \leq T$, and the number of casualties $m$. The counting measure of events associated with this process is denoted $N_X(A \times M) = |X \cap (A \times M)|$ for $A \subseteq W \times [0,T], M \subseteq \mathbb{N}_0$. For any $A \times M$, its \textit{intensity measure} conditional on the history of the process $\hist$ is defined as the expected number of events of $X$ in $A \times M$ given $\hist$, i.e. as $\E(N_X(A \times M)|\hist)$. Its \textit{intensity function} denoted $\lambda(\bs,t,m |\hist)$ is related to this measure via $\E(N_X(A \times M)|\hist)  = \sum_{m \in M} \int_A \lambda(\bs,t,m | \hist) \; \dif \bs \; \dif t$ \citep{moller04}. 

\subsection{Intensity characterization} 
\label{sub:intensity_characterization}
For a given time $t$ and location $\bs$ in our domain of interest ($A= W \times [0,T]$), the likelihood of an event with mark $m$ can be expressed through its intensity function, whose general form \citep{reinhart17} is 
\begin{align}
	\lambda(\bs,t,m ; \hist) & = f_M(m;\Z_m(\bs,t)) \lambda^\ast(\bs,t ; \hist, \Z_{\bs,t}). \label{eq: intensity function}
\end{align}
Here, $f_M(m;\Z_m(\bs,t))$ is a discrete probability density function for the marks given the mark covariates $\Z_m(\bs,t)$ observed at location $\bs$ and time $t$. Further, $\lambda^\ast(\bs,t ; \hist, \Z_{\bs,t})$ is the conditional intensity function of the spatio-temporal process at time $t \in [0, T]$ and spatial location $\bs \in W$ given the history of events $\hist$ and spatio-temporal covariates $\Z_{\bs,t}$.

\subsection{Mark distribution}
Prior data analysis studying continuous extreme valued distributions (GPD and GEV) to country level datasets provided by the GTD, have shown good fits for the number of casualties \citep{patel2020}. In particular, \cite{patel2020} detail that the GPD or POT approach provides the best overall fit to the tails of casualty data, caveatted by a requirement for an exceedance threshold to be determined and the distribution's prime suitability for continuous data. To deal with this, we follow the work of \cite{RN211} and utilize a discrete \textit{mixture} distribution that is able to model an extreme number of (discrete) casualties produced from each attack and also its zero-inflated behavior, as highlighted by the quantile behavior shown in Figures \ref{fig:emp_cdf}-\ref{fig:quantile_function}.

After fixing an exceedance threshold $u \in \mathbb{N}$, we utilize a zero-inflated discrete distribution for all marks $m \in [0,u]$ \textit{below} a threshold $u$ that supports observations of attacks with a low number of casualties. Above $u$, we consider a discrete extreme value distribution addressing attacks resulting in an extreme number of casualties. We model $f_M(m ; \Z_m(\bs,t))$ via the mixture distribution


\begin{align}
	f_M(m; \pi_M, u, \btheta_{ZI}, \btheta_{EV}) &=\pi_M \frac{f_{ZI}(m; u, \btheta_{ZI})}{\sum_{m' \leq u} f_{ZI}(m'; u, \btheta_{ZI})} \mathbbm{1}_{m\leq u} \nonumber \\
	& + 
(1-\pi_M)\frac{f_{EV}(m; u, \btheta_{EV})}{\sum_{m'>u} f_{EV}(m'; u, \btheta_{EV})} \mathbbm{1}_{m>u}, \label{eq: mark distribution}
\end{align}

where $f_{ZI}(m ; u, \btheta_{ZI},\Z_m(\bs,t) )$ denotes a general zero-inflated distribution to model the number of casualties up to threshold $u$ given parameters $\btheta_{ZI}$ and $f_{EV}(m; u, \btheta_{EV}, \Z_m(\bs,t))$ denotes a general extreme valued distribution to model the number of casualties \textit{above} $u$ given parameters $\btheta_{EV}$. The parameter $\pi_M$ denotes the mixing proportion, or weight, of the distribution corresponding to each case.  For simplicity in notation, we drop the explicit dependence on $\Z_m(\bs,t)$ for both $f_{ZI}$ and $f_{EV}.$ 

\subsubsection{Casualty model selection} 
\label{ssub:model_selection} 
Suitable probability mass functions for $f_{ZI}$ include the zero-inflated Poisson (ZIP) and Negative-Binomial (ZINB) distributions, both of whose parameters $\btheta_{ZI}$ can be written as functions of the exogenic covariates $\Z_m(\bs,t)$. For all $m\leq u$, their probability distributions are given by
\begin{align}
	f_{ZIP}(m; u, \btheta_{ZI}) & = \alpha \mathbbm{1}_{m=0} + (1-\alpha) \frac{\lambda^m \mathrm{e}^{-\beta}}{m!} & \alpha \in [0,1], \; \beta>0 \label{eq: zero inflated poisson}
	 \\
	f_{ZINB}(m; u, \btheta_{ZI}) & = \alpha \mathbbm{1}_{m=0} + (1-\alpha) \binom{m+r-1}{m} (1-p)^r p^m & \alpha,p \in [0,1], \; r>0.
\end{align}

Determining a form for $f_{EV}$ requires considering concepts from \textit{discrete extremal functions} which are suitable to this problem. Specifically, for $m \geq u$, we consider the \textit{Generalized-ZIPF} distribution (GZD) defined by \cite{RN195} as
\begin{align} 
	f_{GZD}(m;\btheta_{EV}) & \propto \begin{cases}
\left( 1 + \frac{\xi (m-u)}{\sigma} \right)^{-\frac{1}{\xi} - 1} & \xi > 0 \\
\exp \left(-\frac{(m-u)}{\sigma} \right) & \xi = 0,
\end{cases}
\end{align}
a flexible class of extreme valued discrete distributions decaying via the Zipf-Mandelbrot power-law \citep{mandelbrot53}. Zipf's law created by \cite{Zipf49} originally studied the distribution of word frequencies in common language and has resulted in a wide range of Zipf-type distributions within different frameworks \citep{Li92}. An advent of its usage in the social sciences, for example, to analyze voting histories at different levels of social systems \citep{lyra03}, population densities in cities \citep{Gabaix99} and fake social media trends \citep{rastogi16} complementing its extreme valued properties, renders the GZD a suitable candidate in modeling extreme numbers of casualties resulting from terror insurgencies. While the power law distribution utilized in \cite{RN591} to quantify the probability of high casualty-producing terror events belongs to a single parameter family, the GZD belongs to a two-parameter family of distributions and can be seen as the discrete counterpart of the Generalized Pareto Distribution (GPD), giving extra flexibility in model fitting and easily incorporating a threshold $u$, beyond which extreme events are modeled.

\begin{figure}[htbp]
	\centering
	\includegraphics[width=.49\textwidth]{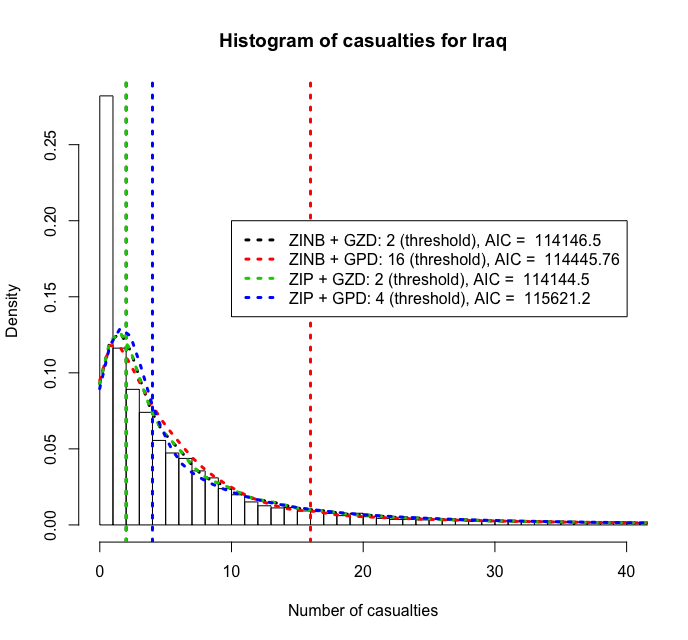}
	\includegraphics[width=.49\textwidth]{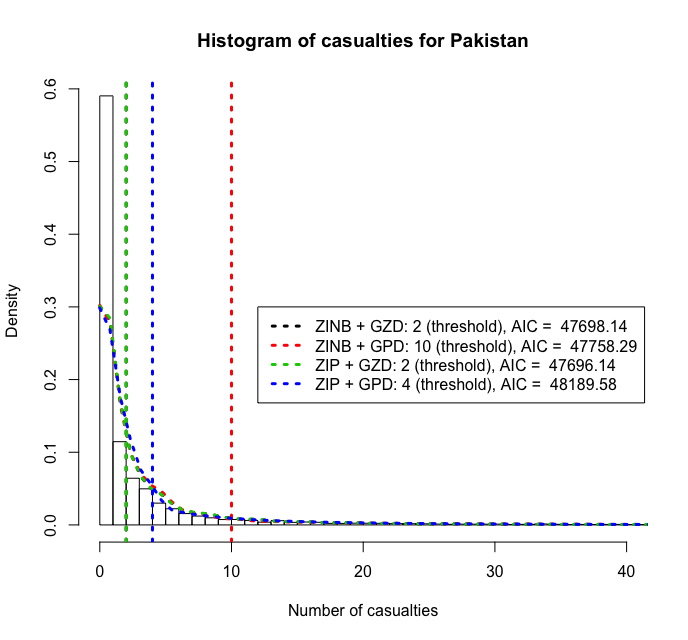} \\
	\includegraphics[width=.49\textwidth]{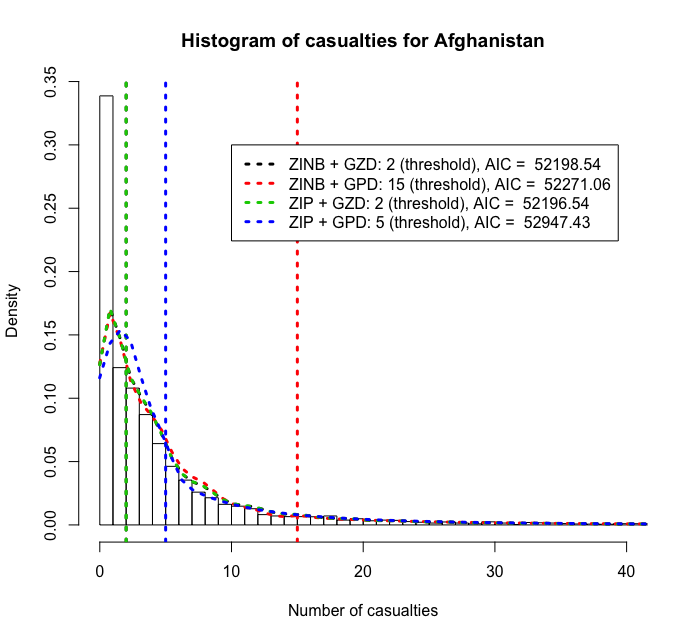}
	\includegraphics[width=.49\textwidth]{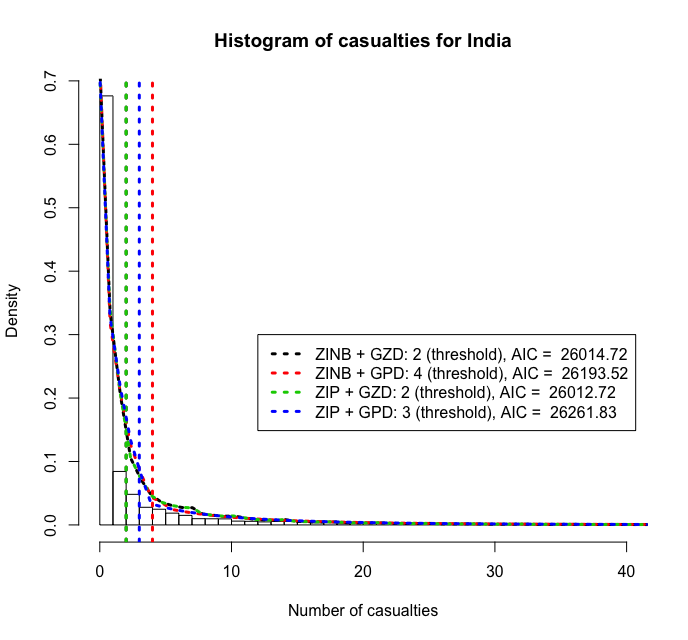}
\caption{Comparisons of mark distributions fit with the GZD/GPD and ZIP/ZINB for Iraq (left, top), Pakistan (right, top), Afghanistan (left, bottom) and India (right, bottom)}
\label{fig:marks}
\end{figure}

We performed preliminary studies considering casualty data from Iraq, Pakistan, Afghanistan, and India with combinations of the zero-inflated Poisson/Negative Binomial distributions for $f_{ZI}$ and the GZD/GPD distributions for $f_{EV}$ in Equation \refeq{eq: mark distribution}. The available data for each country was fitted to our proposed distribution with maximum likelihood estimation for the parameters $\btheta_{ZI}$, $\btheta_{EV}$, $\pi$ and $u$. Figure \ref{fig:marks} shows fits of $f_M$ with either a zero inflated Negative Binomial or a zero-inflated Poisson distributions for $f_{ZI}$ for Iraq, Pakistan, Afghanistan, and India respectively. The threshold $u$ was considered at a few values indicated by vertical dashed lines and the combination of possible distributions/thresholds assessed via the Akaike information criterion (AIC) as reviewed by \citep{cavanaugh19}. Distributional fits measured by AICs, are shown on top of the casualty data histograms of each country. In all cases, AIC favors a ZIP distribution for $f_{ZI}$ and a GZD form for $f_{EV}$, with a threshold of value $u=2$. This is the mark distribution that will be adopted in the remainder of the paper. 


\subsection{Conditional intensity} 
\label{ssub:conditional_intensity}
The conditional intensity $\lambda^\ast(\bs,t ; \hist)$ appearing in Equation (\refeq{eq: intensity function}) describes the instantaneous measure of a terror event occurring at a space-time location $(\bs,t)$ given historic events $\hist$ occurring before time $t$. The self-exciting nature of terror events renders a Hawkes type intensity \cite{hawkes71}, suitable for this conditional measure. The Hawkes process formulation specifically addresses attributes of self-excitation or contagion, accounting for an arbitrary event's contribution to the likelihood of future event occurrence. This flexible model aspect has provided an important basis for its broad utility and has seen application in modeling financial stock crashes and surges \citep{DaFonseca14}, predicting origin times and magnitudes of earthquakes \citep{Ogata88} and propagation of social media events \citep{Rizoiu17}.
To do so, the Hawkes conditional intensity takes the general form \citep{reinhart17}
\begin{align}
	\lambda^\ast(\bs,t ; \hist) & = \mu^\ast(\bs,t;\btheta_{\mu^\ast}) + \sum_{t_i<t} \phi(t-t_i,|\bs-\bs_i|,m_i;\btheta_{\phi}), \label{eq:cond_int}
\end{align}
where $\hist$ denotes the history of the events and $\mu^\ast(\bs,t;\btheta_{\mu^\ast})$ is the baseline spatio-temporal intensity of terror attacks and $\phi(\cdot)$ denotes the \textit{triggering} function, describing the self-exciting intensity of a future event given the space-time locations and marks associated with previous attacks. Depending on the kernel $\phi(\cdot)$, the excitation may be local or have longer term effects in space and/or time \citep{hawkes71}. The terms $\btheta_{\mu^\ast}$ and $\btheta_{\phi}$ denote hyper-parameters associated with the baseline and triggering intensities, respectively. 
 
In the following, we denote an individual event by its measured time, spatial location and casualty mark as $\bx_t := (\bs_{\bx_t}, t, m_{\bx_t})$, and the point pattern of events as $\X = \cup_t \bx_t$: a realization of the \textit{conditional} point process $X(\bs,t;\hist)$. The conditional intensity in Equation (\ref{eq:cond_int}) describes the instantaneous probability of an arbitrary event $\bx_t$ belonging \textit{either} to the background via the intensity contribution of $\mu^\ast$ or being triggered by a background point. 

From this observation, we can consider a Hawkes process as a \textit{branching process} \citep{Hawkes74} of marked spatial events ordered by their time stamps. Here, events are seen to arrive either through immigration or birth. An immigrant corresponds to an event that results from the baseline intensity $\mu^\ast$, whereas events that are caused by self-excitation are birthed descendants from previous events, that are excited via the triggering kernel $\phi$. We note that when $\phi \equiv 0$ almost everywhere, the Hawkes process as defined in Equation (\ref{eq:cond_int}) reduces to an inhomogeneous Poisson process with rate $\mu^\ast(\bs,t;\btheta_{\mu^\ast})$. In this manner, background points form the centers of \textit{inhomogeneous Poisson clusters} of triggered events drawn from the self-exciting kernel $\phi$ with respect to the immigrant event. The interested reader is directed to \cite{Hawkes74} for more information about this Cluster-based Hawkes Process. 

The point pattern $\X$ can be therefore be succinctly written as the set union of the background $\X_{\mu^\ast}$ and triggered pattern $\X_\phi$, written as 
\begin{align*}
	\X & = \X_{\mu^\ast} \cup \X_\phi\\
	\X_\phi & = \cup_{\bx'_t \in \X_{\mu^\ast}} \X_{\bx'_t}, 
\end{align*}
where $\X_{\bx'_t}$ denotes the triggered pattern of inhomogeneous clusters from background point $\bx'_t \in \X_{\mu^\ast}$ at arbitrary time $t \in [0,T]$. 

\subsubsection{Baseline intensity} 
\label{ssub:baseline_intensity}
The baseline intensity $\mu^\ast(\bs,t;\btheta_{\mu^\ast})$ is intended to characterize untriggered terror events, for example, attacks that are initiated by new insurgency groups. In the simplest case, a (constant baseline) homogeneous Poisson intensity i.e. where $\mu^\ast(\bs,t;\btheta_{\mu^\ast}) \equiv \mu^\ast$ describes a process in which the expected number of baseline attacks per unit area is $\mu^\ast$. In general, baseline attacks appear inhomogeneous across areas of interest, and are seen to be correlated with population density, distance to nearest cities and other socio-economic factors \citep{RN597}. To deal with this inhomogeneity and the addition of covariates that can be measured for each region, we model the baseline intensity as an inhomogeneous Poisson process
\begin{align*}
	\mu^\ast(\bs,t;\btheta_{\mu^\ast}) & = \exp(\Z_{\bs,t}^\top \btheta_{\mu^\ast}),
\end{align*}
where $\btheta_{\mu^\ast} \in \mathbb{R}^n$ is a vector of coefficients and $\Z_{\bs,t}$ is a vector of recorded covariates. Covariates of interest include population density, government voting statistics and number of different languages spoken at the district level with GDP, geographical distances to nearest cities and terrains used at the country level. These covariates are available from Statistical Year Books of countries of interest. In the case of Afghanistan, year books are available at \url{https://nsia.gov.af/library}. 
Given a background point pattern $\X_{\mu^\ast}$, the density of the inhomogeneous Poisson intensity (with respect to the unit rate Poisson on $W \times T$) takes the form \citep{baddeley00}
\begin{align}
	f_{\mu^\ast}(\X_{\mu^\ast};\btheta_{\mu^\ast}) = \prod_{\bx'_t \in \X_{\mu^\ast}} \mu^\ast(\bs_{\bx'_t},t;\btheta_{\mu^\ast}) \exp \left(- \int_{W \times T} (\mu^\ast(t',\bs';\btheta_{\mu^\ast})-1) \dif \bs' \dif t' \right), \label{eq:background density}
\end{align}
where the product is over all events with arbitrary time $t \in [0,T]$. Given $\X_{\mu^\ast}$, the point process density in Equation (\ref{eq:background density}) can be numerically maximized to determine $\btheta_{\mu^\ast}$. Here, the integral term can be computed using the quadrature scheme described in \cite{baddeley00} and permits a computationally efficient maximization algorithm. 

\subsubsection{Triggering intensity} 
\label{ssub:triggering_intensity} 
In order to derive the log-likelihood associated with the triggered point pattern $\X_\phi$, it is necessary to understand the baseline origin of such points within the cluster-based Hawkes formulation previously mentioned. In particular, the marginal distribution $p(\X_\phi;\btheta_\phi)$ requires conditioning on all background events in $\X_{\mu^\ast}$ and data sequences $\D_\phi$ from $\X_\phi$ that give time-ordered clusters of triggered events originating from a background point. Following \cite{walder2018}, the log density of $p(\X_\phi;\btheta_\phi)$ is shown to follow
\begin{align}
	\log p(\X_\phi;\btheta_\phi) & = \sum_{d \in \D_\phi} \hspace{-3mm} \sum_{\substack{\bx_{u},\bx_{v} \in d\\ u<v \\ \{d \ni \bx_s: s \in (u,v)\} = \emptyset}} \hspace{-3mm} \sum_{\substack{\bx'_t \in \X_{\mu^\ast} \\ t<u}} \log \phi(\delta(\bx'_t,\bx_{u},\bx_{v})) - \int_{\substack{W \times [0,T-t]}} \phi(t',\bs',m_{\bx'_t}) \dif \bs' \dif t', \label{eq:full_dens_triggering}
\end{align}
where the second summand is over \textit{subsequent} event pairs $\bx_{u},\bx_{v}$ with $v>u$ and $$\delta(\bx'_t,\bx_{u},\bx_{v}) = (v-u,|\bs_{\bx_{v}}-\bs_{\bx_{u}}|,m_{\bx'_t}).$$

This density utilizes the branching structure of the Hawkes' immigration-birth formation through the clusters in $\D_\phi$ as additional variables to consider. Since $\D_{\mu^\ast} = \D \setminus \D_\phi$ can be readily obtained from $\D_\phi$, the self-exciting and background intensity parameters can be evaluated from their derived conditional distributions if this structure is known. 

The triggering function $\phi(\cdot)$ is typically written with a known parametric form. In the one dimensional case,  \cite{hawkes71} proposed the triggering function to be an exponential decay function $\phi(s)  = \alpha \mathrm{e}^{-\beta s}$ with $\alpha,\beta >0$ or power law functions of the form $\phi(s)  = \frac{k}{(c+s)^p}$ for $k,c,p >0$. 

\begin{figure}[htbp]
	\centering
	\includegraphics[width=.33\textwidth]{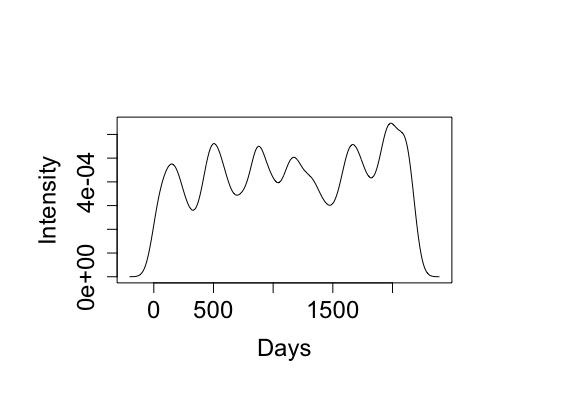}\hfill
	\includegraphics[width=.34\textwidth]{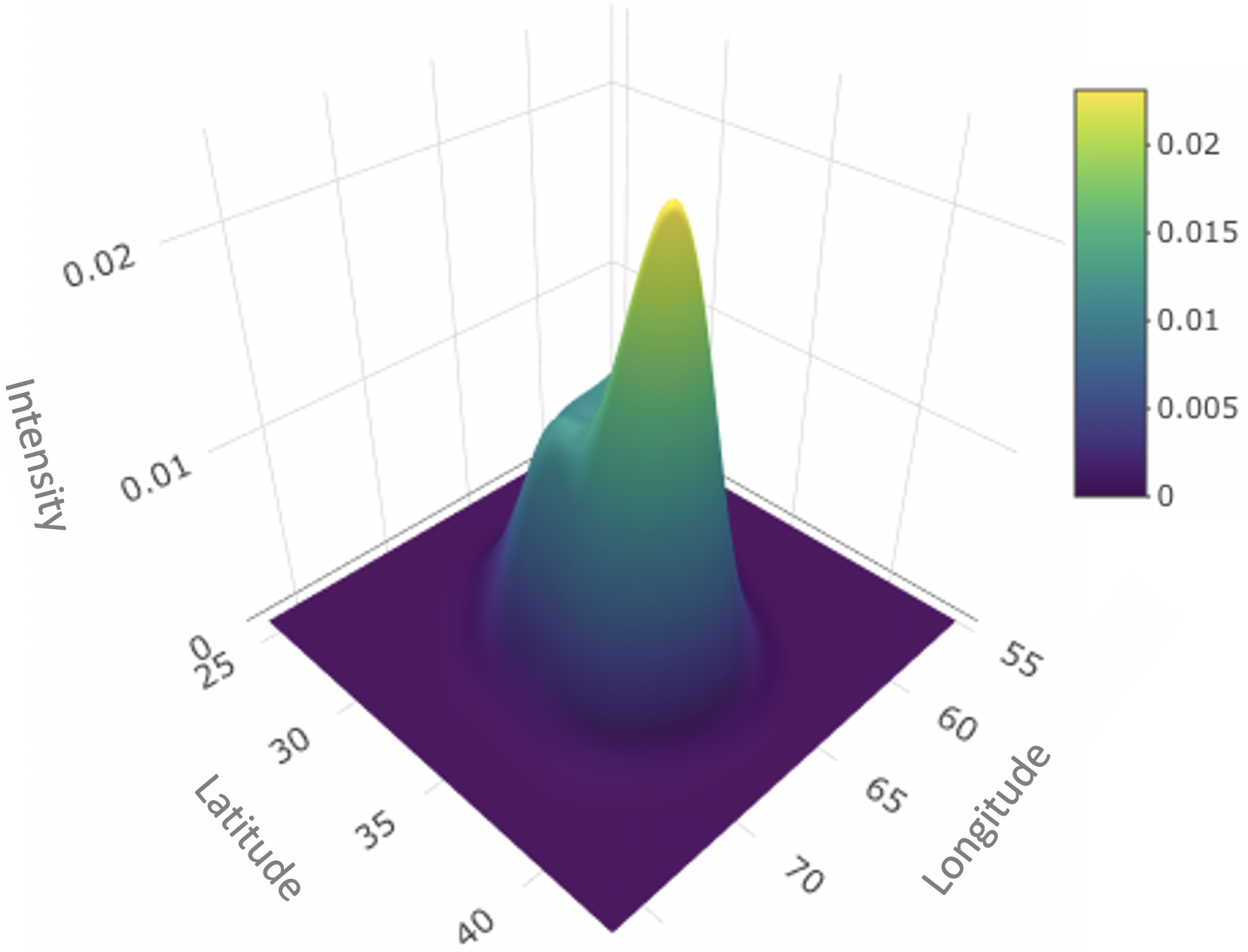}\hfill
	\includegraphics[width=.33\textwidth]{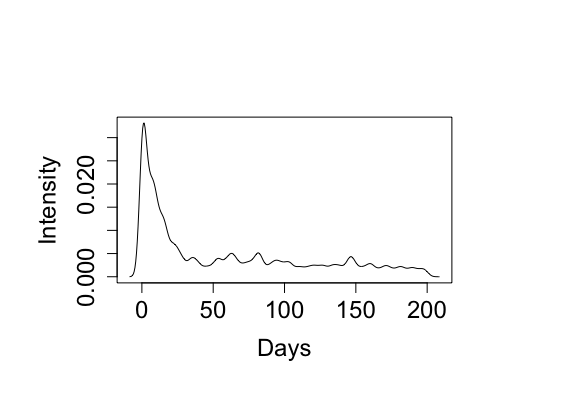}
\caption{Nonparametric Hawkes intensity fits for Afghanistan (2013-2018): Estimated background temporal $\nu_1(t)$ (left), background spatial $\nu_2(\bs)$ (center) and triggering temporal $g_1(\cdot)$ (right) intensities.}
\label{fig:hawkes1}
\end{figure}

For this problem, however, the nuances in terror insurgence excitation and uncertainty surrounding the quantitative effect of historic events render non-parametric models to be more appealing. To show this explicitly, a simplified non-parametric Hawkes process utilized in \cite{RN593}, with intensity 
$$\lambda^\ast(\bs,t ; \mathcal{H}_t) = \nu_1(t)\nu_2(\bs) + \sum_{t_i<t} g_1(t-t_i)g_2(|\bs-\bs_i|),$$
was considered for exploratory data analysis. In relation to the generalized intensity function given in Equation (\ref{eq: intensity function}), we have in this case that $\mu^\ast(\bs,t;\btheta_{\mu^\ast}) = \nu_1(t)\nu_1(\bs)$ and $\phi(t-t_i,|\bs-\bs_i|,m_i) = g_1(t-t_i)g_2(|\bs-\bs_i|)$, where the functions $\nu_1, \nu_2, g_1, g_2$ are to be determined. Empirically based kernel density estimates of the functions $\nu_1, \nu_2$ and $g_1$ fitted to Afghanistan's terror events in the GTD are shown in Figure \ref{fig:hawkes1}. It is seen here that the estimated functions for the temporal background and triggering functions are highly nonlinear and difficult to parametrize. Additionally, this approach cannot easily incorporate additional covariates $\Z_{\bs,t}$ into the procedure, thereby owing to crude spatial intensity estimates.

In order to better understand and characterize terror excitation, the demonstration in Figure \ref{fig:hawkes1} motivates our novel method, presented next, for modeling the triggering function. 

 Specifically, we place a Gaussian Process prior over the class of all triggering functions $\phi(\cdot)$. In a similar manner to the works of \cite{flaxman2017}, the triggering kernel $\phi(t-t_i,|\bs-\bs_i|,m_i)$ is written non-parametrically as $\phi(\cdot) = af^2(\cdot)$, where $f$ is a Gaussian Process and $a>0$, to ensure positivity of the intensity function. The \textit{covariance} function of $f$ can be decomposed into its eigenvalues $\{\tilde{\lambda}_1, \tilde{\lambda}_2, \dots,\}$ and eigenfunctions $\{e_1(\cdot), e_2(\cdot), \dots \}$ via {\it Mercer's theorem} to give 
 $$k(\bx,\by) = \sum_{i=1}^\infty \tilde{\lambda}_i e_i(\bx) e_i(\by).$$ 

The above representation implies that $f(\cdot) = \bomega^\top \mathbf{e}(\cdot)$, where $\bomega \sim \N(0,\tilde{\Lambda} = \text{diag}(\tilde{\lambda}_1, \tilde{\lambda}_2, \dots))$ and $\mathbf{e}(\cdot) = [e_1(\cdot), e_2(\cdot), \dots]$ \citep{zhang2018}. In practice, a low-rank approximation of the above infinite sum is used to ensure computational stability.
			
Typical examples of the covariance function $k(\bx,\by)$ include the {\it Squared exponential} $k(\bx,\by) = \exp \left(-\|\bx-\by\|^2 / 2l^2\right)$ with known $\tilde{\lambda}_i,e_i(\bx)$ with respect to the Gaussian measure, the {\it Rational quadratic}, $k(\bx,\by) = \left(1 + \|\bx-\by\|^2 / 2 \alpha_s l^2\right)^{-\alpha}$ and the {\it Periodic} $k(\bx,\by) = 1 + \sum_{m=1}^\infty 2 \cos(2 \pi m (\bx-\by)) / (2\pi m)^{2s},$ on $[0,1]$ with known $\tilde{\lambda}_i,e_i(\bx)$ with respect to Lebesgue measure \citep{flaxman2017}. Due to this characterization, not all covariance functions have known Mercer expansions.
 
To deal with this, and to allow full flexibility in the choice of a covariance function for $f$, we utilize the work of \cite{flaxman2017}. Specifically, for a positive definite kernel function $k: S \times S \to \mathbb{R}$ over a non-empty domain $S$, a unique reproducing kernel Hilbert space (RKHS) $\mathcal{H}_k$ exists and defines a space of functions $f:S\to\mathbb{R}$ from which $f$ can be represented by an inner product $f(\bx) = \langle f,k(\bx,\cdot) \rangle_{\mathcal{H}_k}$. Using this set-up, for \textit{triggering} data $\bx_i \in \X_\phi$ arising from the event $\bx_j \in \X_{\mu^\ast}$, the function $f$ satisfies 
\begin{align}
	f = \arg\min_{f' \in \mathcal{H}_k} & \left\{ -\sum_i \log(af'^2(t_j-t_i,|\bs_i-\bs_j|,m_i)) + a \int_{W\times T} f'^2(t_j-t,|\bs-\bs_j|,m_i) \; \dif \bs \; \dif t +\gamma\|f'\|^2 \right\}, \label{eq:objective function}
\end{align}
including a regularization term $\gamma$ that corresponds to the squared Hilbert space norm of $f$.
The integral term of \refeq{eq:objective function} does not however guarantee a solution with covariance $k(\bx,\by)$, stemming from the possibility of non-explicit Mercer expansions belonging to the most flexible class of covariances. To deal with this, \cite{flaxman2017} constructs an alternative kernel $\tilde{k}(\bx,\by)$, which can be computed directly if the covariance function has an explicit Mercer expansion with respect to the Lebesgue measure. In general, using the covariance 
$$\tilde{k}(\bx,\by) = \sum_{i=1}^K \frac{\eta_i}{a \eta_i + \gamma} \hat{e}_i(\bx) \hat{e}_i(\by),$$ 
leads to the existence of a solution $f$. 

In the case when Mercer's expansion is unknown, the adjusted eigenvalues $\eta_i / a \eta_i + \gamma$ and eigenvectors $\hat{e}_i$ of kernel $\tilde{k}$ are sampled via the uniform sampling approach described in detail in \cite{flaxman2017} and briefly described here. By scaling $W = (0,1)^2,T = 1,M = (0,1)$, a uniform form grid $\bu$ over all possible values of $\bs, t,m$ can be constructed and the covariance function $K_{\bu \bu}$ computed on the matrix. Implementing {\it Lanczos iterations} invokes the $p$ highest eigenvalues $\tilde{\lambda}_i$ and eigenvectors $e_i^\bu$ to be found of the truncated eigen-decomposition of $K_{\bu \bu}$. Finally, after determining the covariance $K_{\bx \bu}$ between data $\bx$ and $\bu$, $\tilde{k}(\bx,\by)$ can be evaluated explicitly by computing 
$$\eta_i = \frac{\tilde{\lambda}_i}{p}, \qquad \hat{e}_i(\bx) = \frac{\sqrt{p}}{\tilde{\lambda}_i} K_{\bx\bu} e_i^\bu.$$ The above representation thereby allows computation of $\phi(\cdot) = af^2(\cdot)$, where 
\begin{align}
	f(\cdot) & = \bomega^\top \mathbf{\hat{e}(\cdot)}, & \mathbf{\hat{e}(\cdot)} = [\hat{e}_1(\cdot) \dots \hat{e}_p(\cdot)] ,\\
	\bomega & \sim \N \left(0,\tilde{\Lambda} = \text{diag} \left(\frac{\eta_1}{a \eta_1 + \gamma}, \dots, \frac{\eta_p}{a \eta_p + \gamma}\right)\right), & \bomega \in \mathbb{R}^p.\label{eq:omega_dist}
\end{align}
Here, the unknown parameter set $\btheta_\phi = \{\btheta_K, \bomega, a, \gamma\}$, with $\btheta_K$ containing all parameters of the intended covariance structure, is to be determined.

\subsection{MCMC estimation}
Using $\lambda(\bs,t,m ; \hist)$ as given by Equation (\ref{eq: intensity function}), the log-likelihood function of data $\X$ given the unknown parameter sets $\btheta_1 = \{\pi,\btheta_{ZI}, \btheta_{EV}\}$ and $\btheta_2 = \{\btheta_{\mu^\ast}, \btheta_\phi\}$ is 
\begin{align}
	\ell(\btheta_1,\btheta_2;\X) & = \sum_{i=1}^n \left[\log \left(\Z_{\bs,t}^\top \btheta_{\mu^\ast} + \sum_{t_j<t_i} \phi(t_i-t_j,|\bs_i-\bs_j|,; m_i) \right) + \log (f_M(m_i; \btheta_1,\Z_m(\bs,t))) \right] \nonumber \\
&  - \int_{W \times T} \lambda^\ast(\bs,t|\mathcal{H}_t,\Z_{\bs,t},\btheta_2) \; 
\mathrm{d} \bs \; \mathrm{d} t. \label{eq:cond_int_loglik}
\end{align}
Since the intensity function comes in a separable form between the mark and spatio-temporal parameters, we can estimate $\btheta_1$ and $\btheta_2$ in parallel. The parameter set $\btheta_2$ can either be estimated through maximum likelihood methods or via a Metropolis Hastings MCMC algorithm (MCMC). We elect here to use MCMC which samples $\btheta_1$ from the mark vector $\bm = \left(m_1,m_2, \dots, m_n \right)$ via
\begin{align*}
	p(\btheta_1;\bm,\Z_\bm(\bs,t)) & \propto \pi(\btheta_1) \prod_{i=1}^n f_M(m_i;\btheta_1,\Z_{m_i}(\bs,t)),
\end{align*}
with the prior specifications
\begin{align*}
	\pi(\btheta_\beta), \pi(\btheta_\xi), \pi(\btheta_\sigma) & \sim \N(0,I) \\
	\log \left(\frac{1-\alpha}{\alpha}\right), \; \log \left(\frac{1-\pi_M}{\pi_M}\right) & \sim \N(0,1).
\end{align*}

We perform estimation of $\btheta_2$ within a Bayesian Hierarchical framework using a hybrid Metropolis-within-Gibbs particle MCMC algorithm. This utilizes the aforementioned branching structure of the Hawkes' immigration-birth formation as an additional variable, from which both self-exciting and background intensity parameters can be updated using the conditional distributions we will henceforth derive. 

Specifically, the branching variable, denoted $\B = \{C_i\}_i$ consists of inhomogeneous Poisson clusters containing information of all \textit{independent} events $\X_{\mu^\ast} \in \X$ immigrating from the background intensity function and the triggered events $\X_\phi \in \X \setminus \X_{\mu^\ast}$ that are birthed from them. For example, having cluster $C_1 = (\bx_{t_0},\bx_{t_1},\bx_{t_2})$ where $t_0 < t_1 < t_2$, would imply that background event $\bx_{t_0}$ triggers subsequent events $\bx_{t_1}$ and $\bx_{t_2}$ and that this structure belongs to $\B$. From herein, we denote the background event triggering arbitrary cluster $C$ as $C_0 \in \X_{\mu^\ast}$ and the births as $C_{>0}$. 

In order to derive Gibbs conditional distributions parameters in $\btheta_2$, the branching structure $\B$ must be sampled from. Sampling $\B$ from its posterior $p(\B;\X,\btheta_2)$ consists of computing probabilities that an arbitrary event $\bx_{t_i}$ is either an immigrant or birth. To do so, we define the probabilities $p_{ij}$ for all $t_j<t_i$ as the probability event $\bx_{t_i}$ is triggered by event $\bx_{t_j}$ and $p_{i0}$ as the probability event $\bx_{t_i}$ is birthed from the background. Using the intensity in Equation (\ref{eq: intensity function}), we have that 
\begin{align}
	p_{i0} & = \mu^\ast(\bs_i,t_i;\btheta_{\mu^\ast})\left(\mu^\ast(\bs_i,t_i;\btheta_{\mu^\ast}) + \sum_{t_k<t_i} \phi(t_i-t_k,|\bs_i-\bs_k|,m_i) \right)^{-1} \\
	p_{ij} & = \phi(t_i-t_j,|\bs_i-\bs_j|,m_i)\left(\mu^\ast(\bs_i,t_i;\btheta_{\mu^\ast}) + \sum_{t_k<t_i} \phi(t_i-t_k,|\bs_i-\bs_k|,m_i) \right)^{-1} \; \; t_j<t_i. 
\end{align}
These probabilities can be determined in each MCMC iteration given the current value of $\btheta_2$ thereby enabling the branching structure $\B$ to be sampled by sampling the clusters from these probability vectors. 

Once $\B$ is updated, the algorithm then proceeds to sample from the background posterior $p(\btheta_{\mu^\ast};\X,\B,\btheta_2)$ and self-exciting posterior $p(\btheta_\phi;\X,\B,\btheta_2)$. 

For the background density, we note that using the likelihood function given by Equation ($\ref{eq:background density}$), we may update $\btheta_{\mu^\ast}$ using a Metropolis-Hastings step which samples from 
\begin{align*}
	p(\btheta_{\mu^\ast};\X,\B,\Z_{\bs,t}) & \propto f_{\mu^\ast}(\X_{\mu^\ast}|\btheta_{\mu^\ast}) \pi(\btheta_{\mu^\ast}) \\
	\pi(\btheta_{\mu^\ast}) & \sim \N(0,I), 
\end{align*}
where $\X_{\mu^\ast}$ can be determined from $\X$ using $\B$. We note that Gaussian priors are used where applicable for controlling the efficacy of the subsequent MCMC algorithm(s) in accepting a theoretically optimal rate of samples. 

We note here that in the case of a \textit{constant} intensity $\theta_{\mu^\ast}$ resulting from a single constant background covariate, that a Poisson-Gamma conjugacy structure may be used. Specifically, since $|\X_{\mu^\ast}| \sim \text{Poisson}(\theta_{\mu^\ast})$ can be determined from $\B$, we place 
\begin{align*}
p(\theta_{\mu^\ast};\X,\B, \btheta_\phi) & \propto p(\X_{\mu^\ast};\theta_{\mu^\ast}, \btheta_\phi) \pi(\theta_{\mu^\ast}) \\
\pi(\theta_{\mu^\ast}) & \sim \text{Gamma}(1,1) \\ 
\pi(\theta_{\mu^\ast}; |\X_{\mu^\ast}|) & \sim \text{Gamma}(|\X_{\mu^\ast}|+1, 2|W \times T|).
\end{align*} 
For the triggering parameters, we note that given $\B$, the log density given in Equation (\ref{eq:full_dens_triggering}) can be reduced to
\begin{align}
	\log p(\X_\phi;\btheta_\phi,\X,\B) & = \sum_{C \in \B} \hspace{-4mm} \sum_{\substack{\bx_{u},\bx_{v} \in C_{>0}\\ u<v \\ \{C \ni \bx_s: s \in (u,v)\} = \emptyset}} \hspace{-7mm} \log a(\bomega^\top \mathbf{\hat{e}}(\delta(C_0,\bx_{u},\bx_{v})))^2 - a\int_{\substack{W \times \\ [0,T-t_{C_0}]}} (\bomega^\top \mathbf{\hat{e}}(t',\bs',m_{C_0}))^2 \dif \bs' \dif t'. \label{eq:triggering posterior}
\end{align}
Given this, updates for the fixed parameters in $\btheta_\phi$ can be determined via 
\begin{align*}
	p(\btheta_\phi;\X,\B) & \propto p(\X_\phi;\btheta_\phi,\X,\B) \pi(\btheta_\phi) \\ 
	\bomega|a,\gamma & \sim \N \left(0,\tilde{\Lambda} = \text{diag} \left(\frac{\eta_1}{a \eta_1 + \gamma}, \dots, \frac{\eta_K}{a \eta_K + \gamma}\right)\right) \\ 
	\log(a), \log(\gamma) & \sim \N(0,1). 
\end{align*}

In particular, the log posterior $p(\bomega;\X,\B,\btheta_\phi)$ is given by
\begin{align}
	\log p(\bomega;\X,\B,\btheta_\phi \setminus \bomega) & = -\frac{K}{2} \log 2\pi - \frac{1}{2} \left(\log |\tilde{\Lambda}| + \bomega^\top \tilde{\Lambda}^{-1} \bomega \right) + \nonumber \\
	& \sum_{C \in \B} \hspace{-4mm} \sum_{\substack{\bx_{u},\bx_{v} \in C_{>0}\\ u<v \\ \{C \ni \bx_s: s \in (u,v)\} = \emptyset}} \hspace{-7mm} \log a(\bomega^\top \mathbf{\hat{e}}(\delta(C_0,\bx_{u},\bx_{v})))^2 - a \bomega^\top \left(\int_{\substack{W \times \\ [0,T-t_{C_0}]}} E(t',\bs',m_{C_0}) \dif \bs' \dif t' \right) \bomega, \label{eq:log posterior omega}
\end{align}
where for any $\bs,t,m \in W \times [0,T] \times \mathbb{N}_0$, the outer product of $\mathbf{\hat{e}}(t,\bs,m)$ is defined via $$E(t,\bs,m) = \mathbf{\hat{e}}(t,\bs,m) \mathbf{\hat{e}}^\top(t,\bs,m).$$
We note here that computing the integral term in Equation (\ref{eq:log posterior omega}) within the algorithm corresponds to
\begin{align}
	\int_{\substack{W \times \\ [0,T-t_{C_0}]}} \hspace{-3mm} E(t',\bs',m_{C_0}) \dif \bs' \dif t' & = \int_{\substack{W \times \\ [0,T-t_{C_0}]}} \hspace{-1mm} \mathbf{\hat{e}}(t',\bs',m_{C_0}) \mathbf{\hat{e}}(t',\bs',m_{C_0})^\top \dif \bs' \dif t' \nonumber \\
	& = \int_{\substack{W \times \\ [0,T-t_{C_0}]}} \hspace{-1mm} \left [\frac{\sqrt{p}}{\tilde{\lambda}_1} K_{\bx \bu} e_1^\bu \cdots \frac{\sqrt{p}}{\lambda_p} K_{\bx \bu} e_p^\bu \right]  \left[\frac{\sqrt{p}}{\tilde{\lambda}_1} K_{\bx \bu} e_1^\bu \cdots \frac{\sqrt{p}}{\lambda_p} K_{\bx \bu} e_p^\bu \right]^\top \dif \bs' \dif t' \nonumber \\
	& \approx \frac{1}{M} \sum_{i=1}^M \left [\frac{\sqrt{p}}{\tilde{\lambda}_1} K_{\bx'_i \bu} e_1^\bu \cdots \frac{\sqrt{p}}{\lambda_p} K_{\bx'_i \bu} e_p^\bu \right]  \left[\frac{\sqrt{p}}{\tilde{\lambda}_1} K_{\bx'_i \bu} e_1^\bu \cdots \frac{\sqrt{p}}{\lambda_p} K_{\bx'_i \bu} e_p^\bu \right]^\top, \label{eq: importance intergal}
\end{align}
where the $M$ importance samples or particles satisfy $\bx'_i = (t_i(1-t_{C_0}),|\bs_{C_0}-\bs_i|,m_{C_0})$, $\bs_i,t_i$ are uniformly distributed on $W \times [0,T]$ and $\bu$ is a fixed uniform grid in each iteration. Although Equation (\ref{eq: importance intergal}) denotes an approximation to the true value of the integral, we note that using a renewed particle set $\{\bx'_i\}_{i=1}^M$ and even a renewed uniform grid $\bu$ at each iteration will yield an \textit{unbiased} estimate of the true integral, for any $M$. Within an MCMC framework, it is sufficient to compute an unbiased estimate of the log posterior(s) of interest \citep{andrieu10} when computing Metropolis-Hastings acceptance ratios, resulting in matching MCMC convergence properties while limiting the potentially large particle sample size usually required to achieve a favorable approximation. 

Given the varying covariance structures that can be placed on the Gaussian process prior and the unbiased update form for $\phi(\cdot)$, our approach updates the parameter set $\btheta_\phi$ with a Metropolis-Hastings kernel. However, up to an arbitrary rank $p$, studying the posterior distribution of Gaussian process parameter $\bomega$ within this framework is challenging due to potentially high dimensions and variability resulting from the modeling of such nuanced datasets. On the other hand, the ability to differentiate (\ref{eq:log posterior omega}) motivates an approach for updating $\bomega$ using a Hamiltonian Monte Carlo (HMC) step, used extensively in the literature to sample from challenging target distributions \citep{neal2011,betancourt2018}. Specifically, by defining $U(\bomega) = - \log p(\bomega;\X,\B,\btheta_\phi \setminus \bomega)$ termed as the \textit{potential energy} and the momentum vector $\brho \sim \mathcal{N}(0,\Sigma)$ providing the \textit{kinetic energy} $-\log p(\brho;\Sigma)$, HMC is used to jointly sample $\bomega$ and $\brho$ using the \textit{Hamiltonian} or the joint negative log posterior given by
$$H(\brho,\bomega) = U(\bomega) - \frac{1}{2} \brho^\top \Sigma^{-1} \brho.$$
The joint system $(\bomega, \rho)$ consisting of current parameter vector $\bomega$ and momentum $\brho$, initially sampled from its Gaussian prior at each iteration, is evolved via numerical integration of the resulting Hamilton system of equations \citep{betancourt2018}. This requires computation of $\frac{\partial{U(\bomega)}}{\partial{\bomega}}$, which can be given by direct differentiation of Equation (\ref{eq:log posterior omega}) as
\begin{align}
	\frac{\partial{U(\bomega)}}{\partial{\bomega}} & = \tilde{\Lambda}^{-1} \bomega + 2 \sum_{C \in \B} \hspace{-5mm} \sum_{\substack{\bx_{u},\bx_{v} \in C_{>0}\\ u<v \\ \{C_{>0} \ni \bx_s: s \in (u,v)\} = \emptyset}} \hspace{-10mm} & \left[\left(a\int_{\substack{W \times \\ [0,T-t_{C_0}]}} E(t',\bs',m_{C_0})  \dif \bs' \dif t' \right) \bomega - \frac{\mathbf{\hat{e}}(\delta(C_0,\bx_{u},\bx_{v}))}{\bomega^\top \mathbf{\hat{e}}(\delta(C_0,\bx_{u},\bx_{v}))}\right]. \label{eq:derivative log posterior omega}
\end{align}
Once solved to find $(\bomega^\ast, \rho^\ast)$, the solution is accepted with (Metropolis-Hastings') probability $\min(1,\exp(H(\brho,\btheta)-H(\brho^\ast,\btheta^\ast)))$ to account for numerical integration errors that are typically encountered in practice.

\section{Results}\label{sec:results}
We first validate our inference method on simulated data to demonstrate its accuracy in estimating all unknown parameters of the model. We then apply the method to study and predict terror insurgencies in Afghanistan based upon event data detailed in the Global Terrorism Database between 2013 and 2018.

\subsection{Simulated study} 
\label{sub:simulated_study}
To study the accuracy of the developed MCMC inferential algorithm on the conditional intensity estimation, we provide parameter estimates from a simulated example. In this study, 768 events were simulated on the unit hypercube from the self-exciting point process with constant background rate $\mu^\ast = 50$ and with triggering prior covariance
\begin{align}
    K(\bx,\by) & = \exp \left(-\frac{|t_\bx-t_\by|^2}{2l_t^2} \right) \left(1 + \frac{||(\bs_x,m_x) - (\bs_y,m_y)||^2}{2\alpha_s l_s}\right)^{-\alpha_s}, \label{eq:mixed GP cov}
\end{align}
where $\lambda_t = 0.3, \lambda_s = 1, \alpha = 1$ and $a=1, \gamma = 0.1$. The background points were simulated by first generating $\mu' \sim \text{Poi}(\mu^\ast)$ and then uniformly sampling $\mu'$ points over $W \times T \times M$. Second, for each background point $\bx_c := (t_c,\bs_c,m_c),$ inhomogeneous Poisson clusters with intensity $\phi(t-t_c,|\bs-\bs_c|,m_c)$ for $t>t_c$ were generated through rejection sampling that approximated the integrated intensity over the domain via importance weights. Last, the marks were sampled uniformly on $(0,1)$.

\begin{figure}[!t]
    \begin{subfigure}{.33\textwidth}
        \centering
        \includegraphics[scale=.42]{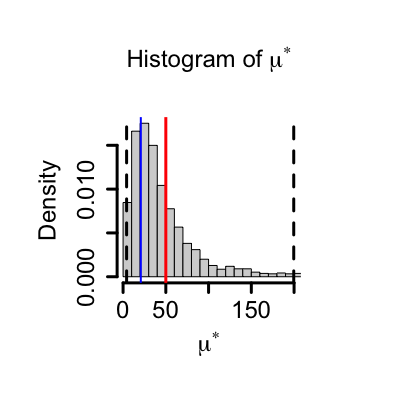}
    \end{subfigure}%
    \begin{subfigure}{.33\textwidth}
        \centering
        \includegraphics[scale=.42]{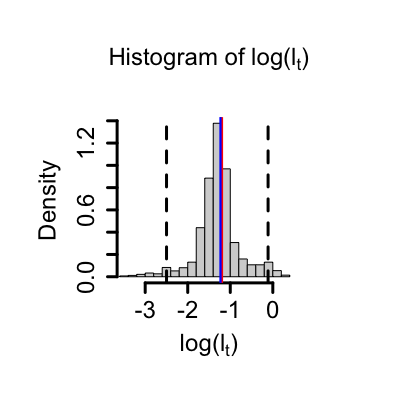}
    \end{subfigure}%
    \begin{subfigure}{.33\textwidth}
        \centering
        \includegraphics[scale=.42]{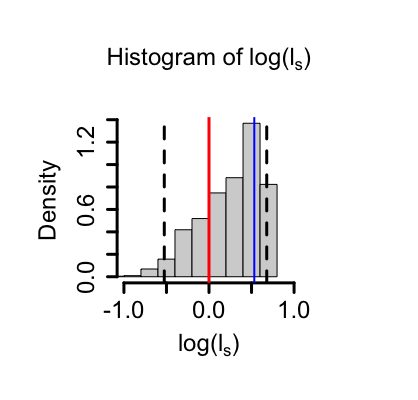}
    \end{subfigure}
        \begin{subfigure}{.33\textwidth}
        \centering
        \includegraphics[scale=.42]{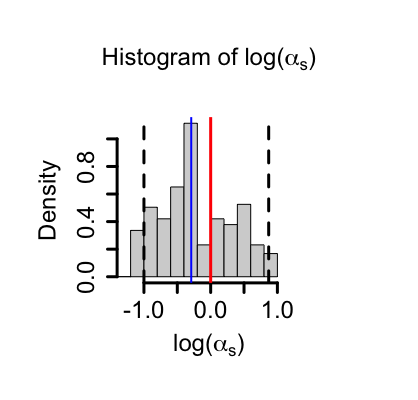}
    \end{subfigure}%
    \begin{subfigure}{.33\textwidth}
        \centering
        \includegraphics[scale=.42]{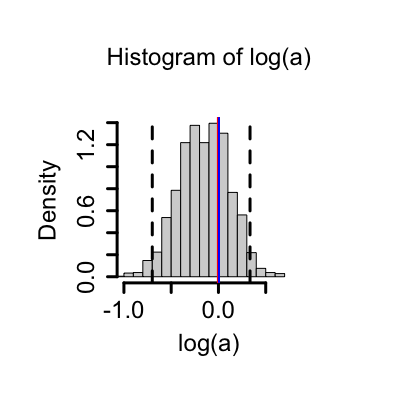}
    \end{subfigure}%
    \begin{subfigure}{.33\textwidth}
        \centering
        \includegraphics[scale=.42]{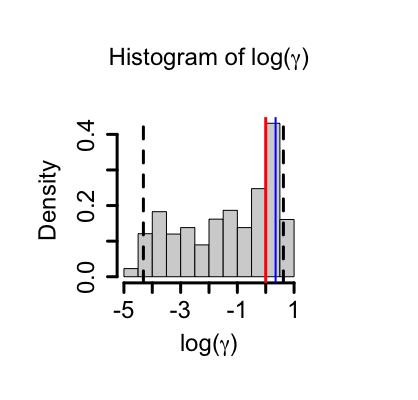}
    \end{subfigure}
    \caption{Posterior plots for all conditional intensity parameters estimated from our algorithm from simulated data. Here, true values (blue) are shown alongside estimated values (red) and 95\% credible intervals (black, dashed). }
\label{fig: post sim}
\end{figure}

Using this methodology, the particle MCMC algorithm was run to obtain $25000$ samples after an initial burn in of $2000$. Posterior parameter estimates for all unknown parameters in the spatio-temporal conditional intensity given in Equation (\ref{eq: intensity function}) were inferred by thinning of the Markov Chain at every tenth sample. Posterior distributions along with posterior modes and 95\% credible intervals for all parameters of interest are given in Figure \ref{fig: post sim}. The true values (blue) and the posterior mode estimates (red) are also shown in the figure.  We notice that the most informative posterior distributions correspond to the parameters $\lambda_t$, $\alpha$ and $a$ also having posterior modes close to the true values. On the other hand, while the posterior mode of $\gamma$ is close to the true value, its distribution shows high variability with heavy tails across its interval. The credible interval for $\mu^\ast$ is much wider and contains its true value, caveatted by the fact that the simulated background rate was estimated from a single realization of a $\text{Poisson}(\mu^\ast)$ variate, and is therefore subject to high posterior variance. It is worth noting that for a relatively small point pattern, all simulated parameters are shown to lie within their 95\% posterior credible intervals. 

\subsection{Modeling Afghan terror (2013-2018)}
We now detail model specificities and results used to portray terror insurgency patterns in Afghanistan between 2013 and 2018. 

\subsubsection{Data preprocessing} 
\label{ssub:data_preprocessing}
The raw event data from the GTD for this study contains some missing values in the form of unknown event times (as days), locations (as latitude/longitude values) and covariates. To deal with this issue, the data used was preprocessed and missing values imputed through the Multivariate Imputation via Chained Equations (MICE) approach \citep{Schafer97} to generate a full dataset. Further, times were jittered uniformly across the day to become a continuous variable needed in the model. Latitude and longitude values were also jittered using a Gaussian distribution centered around their given coordinate with a small standard deviation, thereby adding some randomness to event data determined to the nearest city of true event locations. To be able to use the aforementioned methodology in the computation of the conditional intensity, we standardize the times $t_i \in [0,T]$, locations $\bs_i \in W$ and marks $m_i \in M$ to the unit hypercube and post-scale the resulting spatio-temporal intensity by the area $T \times W$. 

\subsubsection{Model fitting} 
\label{ssub:model_fitting}
In this study, over $10^4$ terror events were collected, data preprocessed and the analysis run with our model. To be able to incorporate terror patterns that show dependence between spatial locations and casualties, an initial MCMC run was used to represent the covariance for $f(\cdot)$ as the space-time separable function given in Equation (\ref{eq:mixed GP cov}), additionally positing a shorter relative time scale for terror excitation. When estimated, however, prohibitively large values for $\alpha_s$ deemed the covariance function 
$$K(\bx,\by) = \exp \left(-\frac{|t_\bx-t_\by|^2}{2l_t^2} \right) \exp \left(\frac{||(\bs_x,m_x) - (\bs_y,m_y)||^2}{2l_s^2}\right),$$
more suitable to be used for this dataset. 

The MCMC algorithm enabling estimation of the conditional intensity and mark parameters was run under these specifications. For both MCMC algorithms, $50000$ samples were computed after an initial burn in of $2000$ and parameters inferred by thinning of the Markov Chain at every tenth sample. 

To account for the dependency of parameters $\btheta_{ZI}$ and $\btheta_{EV}$ on potential mark covariates $\Z_m(s,t)$, we write the parameters of $f_M$ as functions of $\Z_m(\bs,t)$ in the following way
\begin{align}
    \beta(\bs,t) & = \exp(\Z_m(\bs,t)^\top \btheta_\beta)  & \xi(\bs,t) & = \exp(\Z_m(\bs,t)^\top \btheta_\xi) & \sigma(\bs,t) & = \exp(\Z_m(\bs,t)^\top \btheta_\sigma), \label{eq: eva parameter forms}
\end{align}
where  $\btheta_{ZI} = \{\alpha, \btheta_\beta\}$ and $\btheta_{EV} = \{\btheta_\xi, \btheta_\sigma\}$.

For the mark distribution given by Equation (\ref{eq: mark distribution}), the hyper-parameters $\btheta_{ZI}$ and $\btheta_{EV}$ were estimated using the forms of the parameters given in Equation (\ref{eq: eva parameter forms}). Here, the general vector of covariates is chosen as $$\Z_m(\bs,t)^\top = \left( \begin{array}{ccc}
1 & t & \text{pop}_{\bc_\bs,\tilde{y}} \exp(-a\|\bs-\bc_\bs\|_2) \end{array}\right),$$
where $\exp(-a\|\bs-\bc_\bs\|_2)$ denotes an inverse distance measure from spatial point $\bs$ to its nearest city $\bc_\bs$, $a \in \mathbb{R}$ and $\text{pop}_{\bc_\bs,\tilde{y}}$ is the raw population estimate at $\bc_\bs$ in year $\tilde{y}$, utilized to inform the number of casualties occurring within proximity to denser populated areas.

While all three parameter sets $\btheta_\beta,\btheta_\xi$ and $\btheta_\sigma$ were tested using this covariate form, anticipated over-parameterizations invoked the Deviance information criterion (DIC) to be computed for parameter combinations that excluded certain parameters conveying posterior modes close to zero. By defining the \textit{deviance} function as \citep{spiegelhalter02} $$D(\btheta_1) = -2 \log p(\btheta_1;\bm, \Z_\bm(\bs,t)),$$ the DIC is given by
$$DIC = 2 \overline{D(\btheta_1)} - D(\hat{\btheta}_1),$$
where $\hat{\btheta}_1$ denotes the modal value of the parameter vector $\btheta_1$ and $\overline{D(\cdot)}$ denotes the mean deviance. This criterion is particularly useful since the first expression can be easily calculated as the average deviance over the obtained MCMC samples, and the second as the deviance evaluated at their corresponding MAP estimates. Similarly to the well known Bayesian Information Criterion (BIC), the model with the smallest DIC value is favored. 
Using this criterion, we found the most suitable model regulating the mark distribution, as shown in Equations (\ref{eq: mark distribution}) and (\ref{eq: eva parameter forms}), is parametrized via 
\begin{align}
\beta(\bs,t) & = \exp(\btheta_{\beta,1} + \btheta_{\beta,t} t + \btheta_{\beta,c} \text{pop}_{\bc,\tilde{y}}\exp(-a_{\beta}\|\bs-\bc_\bs\|_2)) \label{eq:lambda_eva}\\ 
\xi(\bs,t) & = \exp(\btheta_{\xi,1} + \btheta_{\xi,t} t) \label{eq:xi_eva}\\
\sigma(\bs,t) & = \exp(\btheta_{\sigma,1} + \btheta_{\sigma,t} t + \btheta_{\sigma,c} \text{pop}_{\bc,\tilde{y}} \exp(-a_{\sigma}\|\bs-\bc_\bs\|_2)). \label{eq:sigma_eva}
\end{align}
For casualties over the space-time unit hypercube, this parameterization resulted in the posterior estimates and 95\% credible intervals presented in Table \ref{table: eva values}.
\begin{table}[!ht]
\begin{center}
\begin{tabular}{@{}rrrcrrr@{}} \toprule
        \textbf{Parameter} & \textbf{Estimate} & \textbf{95\% CI} && \textbf{Parameter} & \textbf{Estimate} & \textbf{95\% CI} \\ \midrule
        $\pi_M$ & $0.415$ & $(0.406,0.425)$ && $\btheta_{\xi,1}$ & $-0.744$ & $(-0.904,-0.570)$ \\ 
        $\alpha$ & $0.131$ & $(0.099,0.149)$ && $\btheta_{\xi,t}$ & $0.311$ & $(-0.136,0.712)$  \\ 
        $\btheta_{\beta,1}$ & $0.648$ & $(0.542,0.774)$ && $\btheta_{\sigma,1}$ & $1.233$ & $(1.150,1.319)$ \\
        $\btheta_{\beta,t}$ & $-0.539$ & $(-0.815,-0.256)$ && $\btheta_{\sigma,t}$ & $0.953$ & $(0.763,1.180)$ \\
        $\btheta_{\beta,c}$ & $-0.507$ & $(-0.687,-0.324)$ && $\btheta_{\sigma,c}$  & $0.202$ & $(0.066,0.368)$ \\
        $a_\beta$ & $-0.126$ & $(-1.741,1.971)$ && $a_\sigma$  & $0.393$ & $(-1.728,2.033)$ \\  
        \bottomrule 
    \end{tabular}
\end{center}
\caption{Posterior estimates and 95\% credible intervals for parameters in $\beta(\bs,t),\xi(\bs,t)$ and $\sigma(\bs,t)$ (see Equations (\ref{eq:lambda_eva}-\ref{eq:sigma_eva})) of the mixed mark distribution given in Equation (\ref{eq: mark distribution}).}
\label{table: eva values}
\end{table}

For the self-exciting model, the baseline intensity is assumed to have the form 
$\mu^\ast(\bs,t;\btheta_{\mu^\ast}) = \exp(\Z_{\bs,t}^\top \btheta_{\mu^\ast}),$
where the vector of covariates $$\Z_{\bs,t}^\top = \left( \begin{array}{cccccccccc}
1 & x_\bs & y_\bs & x_\bs^2 & y_\bs^2 & t & \text{Pop.dens}_{\bs,t} & \text{Alt}_\bs & \text{Lang}_\bs & \text{Elect.opp}_{\bs,t} \end{array} \right),$$
includes a log-quadratic contribution of the locations to add extra spatial curvature to the resulting intensity. Further, population density $\text{Pop.dens}_{\bs,t}$, altitude $\text{Alt}_\bs$, number of languages spoken $\text{Lang}_\bs$ and the density of population supporting the opposing government $\text{Elect.opp}_{\bs,t}$, at location $\bs$ and time $t$ (where applicable) are included as measurable variables affecting the rate of insurgency. Population and altitude covariates have been used successfully to study events detailed in the Afghan war dairy \citep{RN597}, and here extended to account for continuous spatial variables by exploiting the absolute distance to the nearest city within a province where the covariable has been measured. Further, language distribution and density of population supporting the opposing government were added to account for the effects of ethnic diversity and Taliban repression on civilians to control the country's governing \citep{Tuerk19}. Other covariates including density of civil workers, yearly district revenue and population density in support of the governing party that were initially tested, invoked poor fits due to missing values and were therefore subsequently removed from the model. Unknown covariates, including those measured \textit{yearly} such as population densities, were linearly interpolated to create a full dataset of values over continuous time. Covariates were found from \url{https://nsia.gov.af/library} and \url{https://www.iec.org.af/results/en/home}. 

While the baseline kernel parameters $\btheta_{\mu^\ast}$ were updated using a standard Metropolis-Hastings with a Gaussian proposal adjusted to conceive an approximate 23\% acceptance rate, sampling the triggering parameters $\btheta_{\phi} \setminus \bomega$ in a similar fashion resulted in poor mixing due to high correlations induced by the Gaussian process formulation. Further, formulating a similar HMC algorithm used to update $\bomega$ anticipated analytical challenges due to heavy computations of the log likelihood derivative. To deal with this, we implemented the Adaptive Metropolis (AM) algorithm which is able to exploit the inherent correlation structure between the parameters of $\btheta_\phi \setminus \bomega$. This algorithm constructs a random walk proposal centered at the previous sampled value and using the empirical covariance of previously accepted samples. A further description of this sampler, including a proof detailing its ergodicity, can be found in \cite{haario01}.

\begin{figure}[!t]
\centering
\begin{subfigure}{.49\textwidth}
    \centering
    \includegraphics[scale=.45]{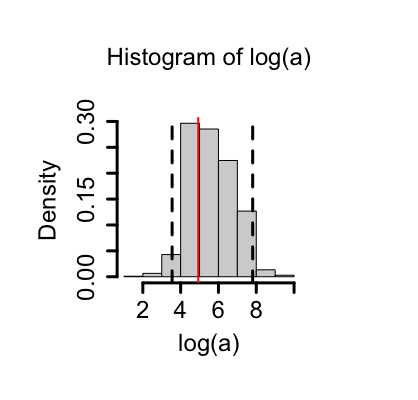}
\end{subfigure}%
\begin{subfigure}{.49\textwidth}
    \centering
    \includegraphics[scale=.45]{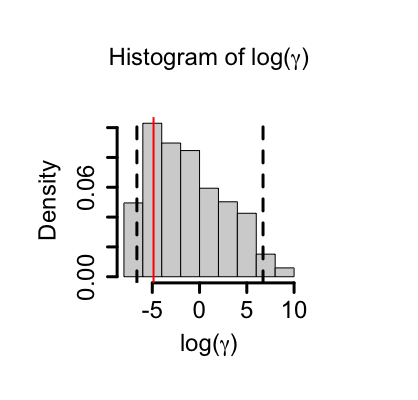}
\end{subfigure} 
\begin{subfigure}{.49\textwidth}
    \centering
    \includegraphics[scale=.45]{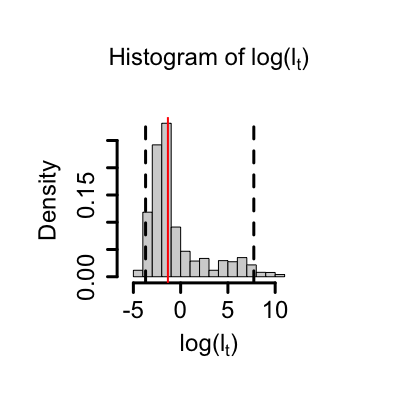}
\end{subfigure}%
\begin{subfigure}{.49\textwidth}
    \centering
    \includegraphics[scale=.45]{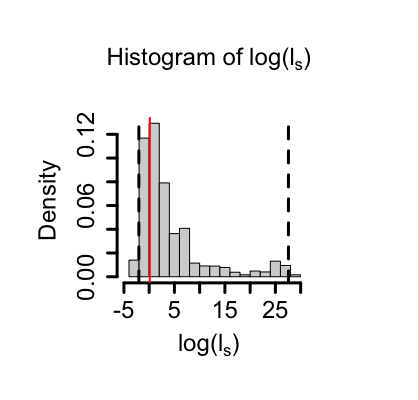}
\end{subfigure}
\caption{Posterior density plots of the triggering intensity parameters $\btheta_\phi$, with the posterior modes (red) shown with 95\% credible intervals (black).}
\label{fig: posterior triggering}
\end{figure}

For marked spatio-temporal events over the unit space-time hypercube, inference of $\btheta_{\phi}$ utilizing the AM algorithm resulted in the posterior densities presented in Figure \ref{fig: posterior triggering}. The correlation between $l_t$ and $l_s$ appearing in the Gaussian process covariance $K(\bx,\by)$ utilized for this dataset is shown by their similarly shaped heavy-tailed posteriors, also highlighting nuanced spatio-temporal insurgency patterns in Afghanistan between 2013-2018. On the other hand, the lower posterior mode of $\log(a)$ in comparison to $\btheta_{\mu^\ast}$ emphasizes the importance of the baseline intensity and its regulating covariables in measuring insurgent events, primarily arising in densely populated areas and districts with low civilian support for insurgent repression. Nonetheless, the relatively high magnitude of triggering events reflected by $\log(a)$ confirms succinct self-excitation patterns across varying scales in space-time, reflecting organized terror events by such groups invoking violence for insurgency success. For the conditional intensity, posterior estimates and 95\% credible intervals of $\btheta_{\mu^\ast}$ and $\btheta_{\phi}$ are presented in Table \ref{table: cov values}. 

\begin{table}[!ht]
\begin{center}
    \begin{tabular}{@{}rrrcrrr@{}}
        \toprule
        \textbf{Parameter} & \textbf{Estimate} & \textbf{95\% CI} && \textbf{Parameter} & \textbf{Estimate} & \textbf{95\% CI} \\ 
        \midrule
        $\btheta_{\mu^\ast,1}$ & $8.413$ & $(7.284,8.523)$ && $\log(l_t)$ & $-1.339$ & $(-3.760,7.729)$ \\ 
        $\btheta_{\mu^\ast,x_\bs}$ & $2.830$ & $(2.343156, 3.812476)$ && $\log(l_s)$ & $0.1221$ & $(-2.042,27.586)$ \\
        $\btheta_{\mu^\ast,y_\bs}$ & $8.140$ & $(7.705,10.239)$ && $\log(a)$ & $4.937$ & $(3.55,7.811)$ \\
        $\btheta_{\mu^\ast,x_\bs^2}$ & $-1.862$ & $(-2.712,-1.078)$ && $\log(\gamma)$ & $-5.867$ & $(-6.628, 6.716)$ \\
        $\btheta_{\mu^\ast,y_\bs^2}$ & $-8.315$ & $(-10.536,-7.821)$ && $\bomega_1$ & $-0.0061$ & $(-0.170,0.064)$ \\
        $\btheta_{\mu^\ast,t}$ & $-0.190$ & $(-0.392,-0.037)$ && $\bomega_2$ & $-0.0161$ & $(-0.225,0.059)$ \\
        $\btheta_{\mu^\ast,\text{Pop.dens}_{\bs,t}}$ & $1.235$ & $(1.152, 1.843)$ && $\bomega_3$ & $-0.0068$ & $(-0.094,0.097)$ \\
        $\btheta_{\mu^\ast,\text{Alt}_\bs}$ & $-1.196$ & $(-1.402,-1.094)$ && $\bomega_4$ & $-0.0015$ & $(-0.164,0.021)$ \\
        $\btheta_{\mu^\ast,\text{Lang}_\bs}$ & $0.047$ & $(0.036,0.103)$ && $\bomega_5$ & $-0.0004$ & $(-0.208,0.028)$ \\  
        $\btheta_{\mu^\ast,\text{Elect.opp}_{\bs,t}}$ & $-0.879$ & $(-1.253,-0.802)$ && $\bomega_6$ & $8.71923 \times 10^{-5}$ & $(-0.0193,0.0314)$ \\
        \bottomrule
    \end{tabular}
\end{center}  
\caption{Posterior estimates and 95\% credible intervals for parameters in $\btheta_{\mu^\ast}$ and $\btheta_{\phi}$, appearing in the conditional intensity function log likelihood (see Equation (\ref{eq:cond_int_loglik})), as determined using a hybrid MCMC approach.}
\label{table: cov values}
\end{table}

\begin{figure}[!h]
\centering
\begin{subfigure}{.34\textwidth}
    \centering
    \includegraphics[width=1.3\textwidth,height=1.35\textwidth]{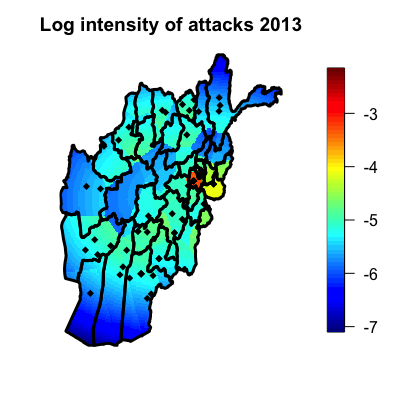}
\end{subfigure}%
\begin{subfigure}{.34\textwidth}
    \centering
    \includegraphics[width=1.3\textwidth,height=1.35\textwidth]{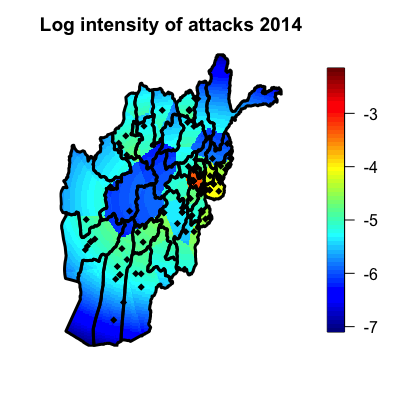}
\end{subfigure}%
\begin{subfigure}{.34\textwidth}
    \centering
    \includegraphics[width=1.3\textwidth,height=1.35\textwidth]{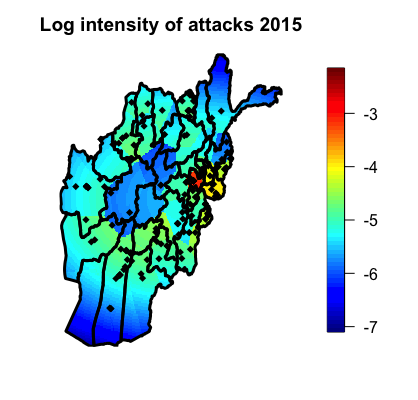}
\end{subfigure}
\begin{subfigure}{.34\textwidth}
    \centering
    \includegraphics[width=1.3\textwidth,height=1.35\textwidth]{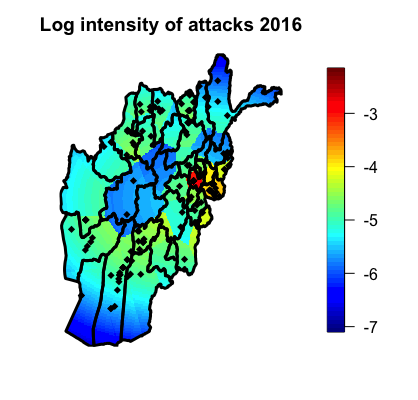}
\end{subfigure}%
\begin{subfigure}{.34\textwidth}
    \centering
    \includegraphics[width=1.3\textwidth,height=1.35\textwidth]{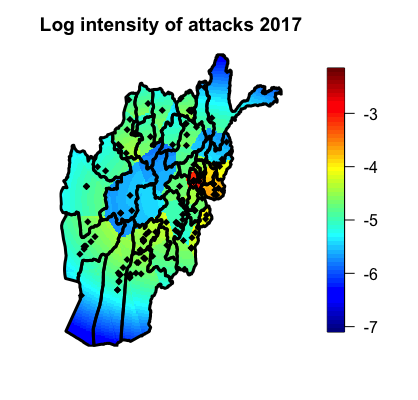}
\end{subfigure}%
\begin{subfigure}{.34\textwidth}
    \centering
    \includegraphics[width=1.3\textwidth,height=1.35\textwidth]{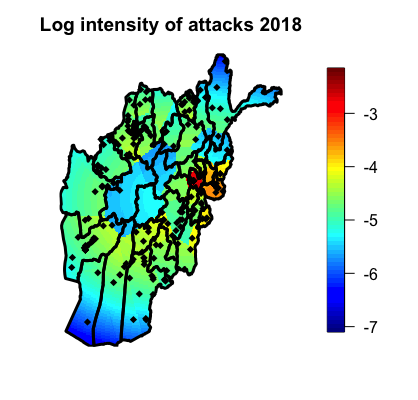}
\end{subfigure}
\caption[short]{Afghanistan spatial-temporal intensity (log-scale) fits for attacks producing at least 20 casualties, between 2013-2018 with events shown as black diamonds.}
\label{fig: afg fits}
\end{figure}

Results using our methodology and posterior parameter estimates are shown in Figure \ref{fig: afg fits}, where the space-time intensity is computed using importance samples of the time variable across the year, therefore presenting average yearly intensities. This intensity is then transformed back to its original units of the observation windows $T \times W \times M$ and shown on the log-scale. Here, the spatio-temporal intensities of attacks leading to more than 20 casualties across the region between 2013 and 2018 are shown, along with recorded events. In line with \citep{RN597}, we see higher intensities along areas targeted by terrorists along the country's ``ring-road'' which visits Kabul, Kandahar, Herat and Balkh, as shown in Figure \ref{fig: afg roads}. An increase in extreme attacks resulting from such activity suggests the gradual insurgence in districts that follow this road. 

\subsection{Extreme Afghan terror prediction 2019-2021}
\begin{figure}[!t]
        \centering
        \includegraphics[width=8cm]{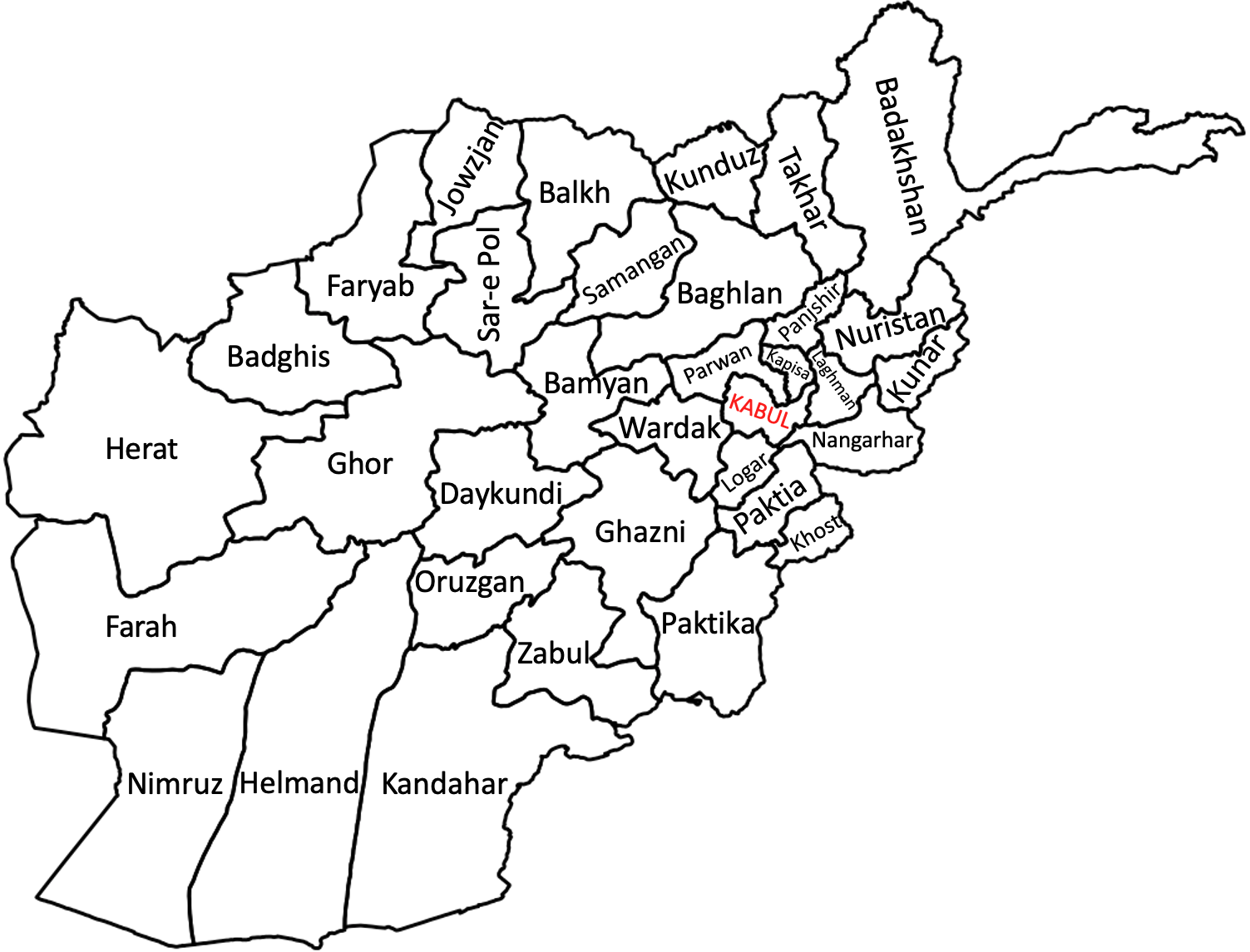}
    \caption{Afghanistan districts by name. Planned new road developments in Afghanistan in 2017 \citep{Amiri2017} highlight districts in which increased extreme terror activity has been predicted by our model (see Figure \ref{fig: afg predictions}).}
\label{fig: afg roads}
\end{figure}

\begin{figure}
\centering
\begin{subfigure}{.34\textwidth}
    \centering
    \includegraphics[width=1.3\textwidth,height=1.35\textwidth]{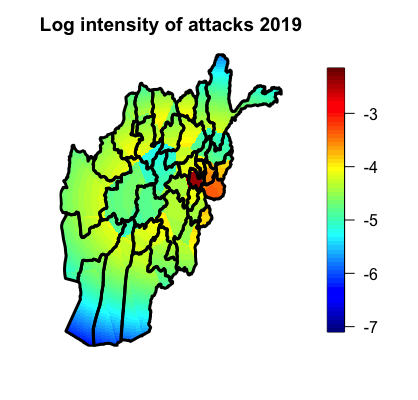}
\end{subfigure}%
\begin{subfigure}{.34\textwidth}
    \centering
    \includegraphics[width=1.3\textwidth,height=1.35\textwidth]{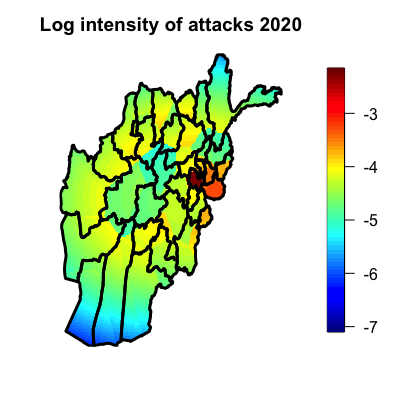}
\end{subfigure}%
\begin{subfigure}{.34\textwidth}
    \centering
    \includegraphics[width=1.3\textwidth,height=1.35\textwidth]{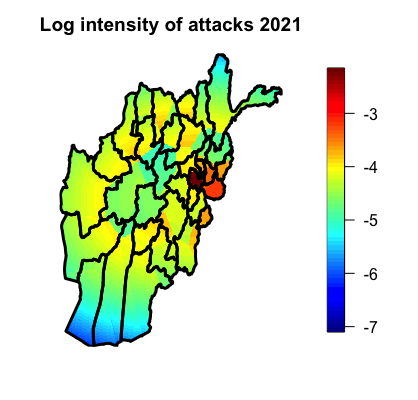}
\end{subfigure}
\caption[short]{Afghanistan spatial-temporal predicted intensity (log-scale) for attacks producing at least 20 casualties, between 2019-2021.}
\label{fig: afg predictions}
\end{figure}

To be able to predict future events in 2019, 2020 and 2021 for which there is no data in the GTD, the time variable $T$ was extended to December 31st 2021. Figure \ref{fig: afg predictions} shows predicted yearly spatio-temporal average intensities of attacks (log-scaled) leading to 20 or more casualties across the region between these years and confirms the fitted intensities for the previous years in a forecasted surge of attacks, leading to the recent insurgency success. This is especially heightened around the ``ring-road'' previously mentioned, with the extremity of insurgent terror around the district of Kabul shown to be increasing, more than doubling the yearly average extreme casualty ($\geq 20$) intensity of 0.001 in 2013 to a predicted of 0.112 in 2021, over pixelated grids of $0.06 \times 0.1$ longitude/latitude units. This increase in attacks also focuses on regions that have not previously seen much terror activity. These areas include the North-eastern Badakhshan Province, a mountainous region which recently has seen increased presence of the Uyghur armed insurgency group, the East Turkestan Islamic Movement \citep{Devonshire2021}. In particular, in this region lies the Wakhan Corridor, connecting Afghanistan to neighboring countries and has been subject to recent road plannings by the Chinese government, as part of the Belt and Road construction \citep{Goulard2021}, for greater market access. Figure \ref{fig: afg roads} shows a disrict map of Afghanistan. \cite{Amiri2017} shows a planned map of new roads to have been developed since 2017, of which districts in the north and center of Afghanistan had been planned for greater infrastructure, aligning closely with an increased risk of terror insurgence predicted by our model. Our analysis therefore emphasizes the increase in extreme terror related activities from new insurgent groups and infrastructure developments. 

\section{Conclusion}\label{sec:disc}
In this paper, we have formulated a flexible spatio-temporal model to study extreme terror insurgencies and study Afghanistan between 2013 and 2021 to show the effectiveness of our model. Our framework importantly focuses on quantifying the risk of \textit{extreme} attacks using a mixed distribution comprising a discrete extreme-valued and a zero inflated model of casualties produced per attack. By incorporating this measure on casualties defined as marks within a spatio-temporal point process framework, the conditional modeling of spatio-temporal trends of extreme insurgencies has been developed. Here, a self-exciting model for terror attacks and resurgences has been constructed through an inhomogeneous Poisson baseline and a Gaussian Process formulated triggering kernel. The former accounts for important covariates related to standard terror behavior, which in the study of Afghanistan includes population density at the country's district level and voting density for the opposing government. The latter develops a flexible prior over all Gaussian Process covariance structures to determine triggering functions that can succinctly incorporate nuanced terror patterns observed in both space and time. All unknown parameters of the model are quantified via a hybrid particle MCMC algorithm which utilizes Mercer's approximation to Gaussian Process covariances. Spatio-temporal prediction of extreme terror insurgencies in Afghanistan using our model, corroborate strongly with recent insurgent events in the country and infrastructure developments that had planned to take place in areas previously unperceived to terror. 

The methodology we have developed provides a basis for several technical extensions. For example, the handling of missing data as studied in \cite{tucker:19} for the GTD could be fully accounted for in the constructed MCMC algorithm, however, is likely to suffer from severe computational drawbacks when analyzing much larger datasets than those studied in this paper. A potential scope for future research would therefore be to investigate avenues providing faster inferential schemes or approximate inference that are compatible with the Gaussian Process methodology we have presented and the missing data problem inherent to the data. Further, differing covariance structures could be studied by our broad methodology to analyze the effect of the unknown triggering function on the overall organization of planned events. 

While this paper focuses on developing a self-exciting point process framework for characterizing spatio-temporal terror patterns of insurgency and their societal risk, the flexibility inherent to our model is likely to be applicable to a broad range of events that are observed in space-time and produce a measurable outcome of interest. For example, stochastic processes underlying stock market return data, extreme weather conditions and cyber-network traffic events could all benefit from our model formulation and inference mechanism to model and predict the risk of extreme events. The novel inclusion of the nonparametric Gaussian Process prior placed upon the triggering kernel particularly highlights the applicability of our work to understudied self-exciting processes that are beyond the scope of this paper. 


\section*{Acknowledgement}
This work was supported by a Sandia National Laboratories Laboratory Directed Research and Development (LDRD) grant. Sandia National Laboratories is a multimission laboratory managed and operated by National Technology and Engineering Solutions of Sandia, LLC., a wholly owned subsidiary of Honeywell International, Inc., for the U.S. Department of Energy's National Nuclear Security Administration under contract DE-NA0003525 and SAND2021-12962 O.  This paper describes objective technical results and analysis. Any subjective views or opinions that might be expressed in the paper do not necessarily represent the views of the U.S. Department of Energy or the United States Government.


\bibliographystyle{chicago}
\bibliography{steem_refs}


\end{document}